\documentclass[twoside]{article}
\usepackage{Proc_NTSE}
\newcommand{\gsim}{\raisebox{-0.7ex}{$\stackrel{\textstyle >}{\sim}$ }}
\newcommand{\lsim}{\raisebox{-0.7ex}{$\stackrel{\textstyle <}{\sim}$ }}

\def\siii{^3 \hskip -0.025in S _1}

\def\ampitrip{7.45^{+0.57}_{-0.53}{}^{+0.71}_{-0.49}}
\def\rmpitrip{3.71^{+0.28}_{-0.31}{}^{+0.28}_{-0.35}}
\def\ampitripphys{1.82^{+0.14}_{-0.13}{}^{+0.17}_{-0.12}}
\def\rmpitripphys{0.906^{+0.068}_{-0.075}{}^{+0.068}_{-0.084}}

\begin{document}
\thispagestyle{plain}

\begin{center}
{\Large \bf \strut
Nuclear Forces from Lattice Quantum Chromodynamics~\footnote{Presentation at the 
{\it International Conference on Nuclear Theory in the Supercomputing Era -
  2013}, Iowa State University, May 13-17, 2013, Ames, Iowa.}
\strut}\\
\vspace{10mm}
{\large \bf 
Martin J. Savage$^{a}$
}
\end{center}

\noindent{
\small $^a$\it Institute for Nuclear Theory, Box 351550, Seattle, WA 98195-1550, USA} 
\\

\markboth{
Martin J. Savage}
{
Nuclear Forces from Lattice QCD
} 

\begin{abstract}
A century of coherent experimental and theoretical investigations have 
uncovered the laws of nature that underly nuclear physics.
The standard model of strong and electroweak interactions, 
with its modest number of input parameters,
dictates the dynamics of the quarks and gluons - the underlying 
building blocks of protons, neutrons, and nuclei.  
While the analytic techniques of quantum field theory have played a key
role in understanding the dynamics of matter in high energy processes, they
encounter difficulties when applied to low-energy nuclear structure and
reactions, and dense systems.
Expected increases in computational resources into the exa-scale 
during the next decade will provide the ability to numerically compute 
a range of important strong interaction processes directly from QCD
with quantifiable uncertainties using the technique of Lattice QCD.
These calculations will refine the chiral nuclear forces that are used as 
input into nuclear many-body calculations, including the three- and 
four-nucleon interactions.
I discuss the state-of-the-art Lattice QCD 
calculations of quantities of interest in nuclear physics,
progress that is expected in the near future, and the impact upon
nuclear physics.
\\[\baselineskip] 
{\bf Keywords:} {\it Nuclear Forces; Lattice QCD}
\end{abstract}

\section{Introduction}

A nucleus is at the heart of every atom, and 
loosely speaking, is a collection of protons and
neutrons that interact pairwise, with much smaller, but
significant, three-body interactions.
We are fortunate to know that  
the underlying laws governing the strong interactions  
result from a quantum field theory called quantum chromodynamics (QCD).
It is constructed in terms of
quark and gluon fields with interactions determined by a local SU(3)
gauge-symmetry and, along with quantum electrodynamics (QED), 
underpins all of nuclear physics when the five relevant input
parameters, the scale of strong interactions $\Lambda_{\rm QCD}$, the three
light-quark masses $m_u$, $m_d$ and $m_s$, and the electromagnetic coupling
$\alpha_e$, are set to their values in nature.
It is remarkable that
the complexity of nuclei
emerges from ``simple'' gauge theories with just five input parameters.
Perhaps even more  remarkable is that  nuclei resemble
collections of nucleons and not collections of quarks and gluons.
By solving QCD, we are expecting to predict, with arbitrary precision, nuclear
processes and the properties of multi-baryon systems.

The fine-tunings 
observed in the structure of nuclei, and in the interactions between nucleons,
are peculiar  
and fascinating aspects of nuclear physics.
For the values of the input parameters that we have in our universe, 
the nucleon-nucleon (NN) interactions are fine-tuned 
to produce unnaturally large scattering lengths in both s-wave channels
(described by 
non-trivial fixed-points in the low-energy effective field theory (EFT)), 
and
the energy levels in the $^8$Be-system, $^{12}$C and $^{16}O$ are in ``just-so''
locations  to produce enough $^{12}$C to support life, and the subsequent
emergence and 
evolution of the human species.  
At a fundamental level it is important for us to determine the sensitivity of the
abundance of $^{12}$C to the light-quark masses and to ascertain the degree of
their fine-tuning.

Being able to solve QCD for the lightest nuclei,
using the numerical technique of Lattice QCD (LQCD), 
would allow for a partial unification of nuclear physics. 
It would be possible to ``match'' the traditional nuclear physics
techniques - the solution of the quantum many-body problem for neutrons and
protons using techniques such as 
No-Core Shell Model (NCSM), Greens function Monte Carlo
(GFMC), and others, to make predictions for the structure and interactions of
nuclei for larger systems than can be directly calculated with LQCD.
By placing these calculations on a fundamental footing, reliable predictions
with quantifiable uncertainties can then be made for larger systems.

\section{Chiral Nuclear Forces}
During the 1990's, the nuclear forces were systematized by the
hierarchy emerging from the spontaneously broken chiral symmetries of QCD.
The resulting small expansion parameters are powers of the external momenta and
powers of the light-quark masses normalized to the scale of chiral symmetry
breaking, as pioneered by
Weinberg, 
first in the meson sector and then the
multi-nucleon sector~~\cite{Weinberg:1990rz,Weinberg:1991um,Weinberg:1992yk}.  
In addition to generating  nuclear forces that are consistent with QCD, this
construction provides the calculational advantage of parametric estimates of
the systematic uncertainty introduced by the truncation of the nuclear
interactions at a  given order in the expansion.
The actual ordering of contributions  remains a subject of debate even today,
with Weinberg's chiral expansion of the potential having its peculiar
difficulties, as does the KSW expansion of scattering amplitudes~\cite{Kaplan:1998tg,Kaplan:1998we}. 
Calculations are being performed at a sufficiently high order where the size of
truncation errors are quite small.  
Weinberg's ordering of operators based upon
a chiral expansion of the n-body potentials between nucleons has been carried
out to N$^3$LO, which includes
contributions to the three-body (starting at N$^2$LO) and the leading 
four-body interactions (starting  at  N$^3$LO)
(for a recent review see Ref.~\cite{Epelbaum:2012vx}).

\begin{figure}[!ht]
\centerline{\includegraphics[width=0.5\textwidth]{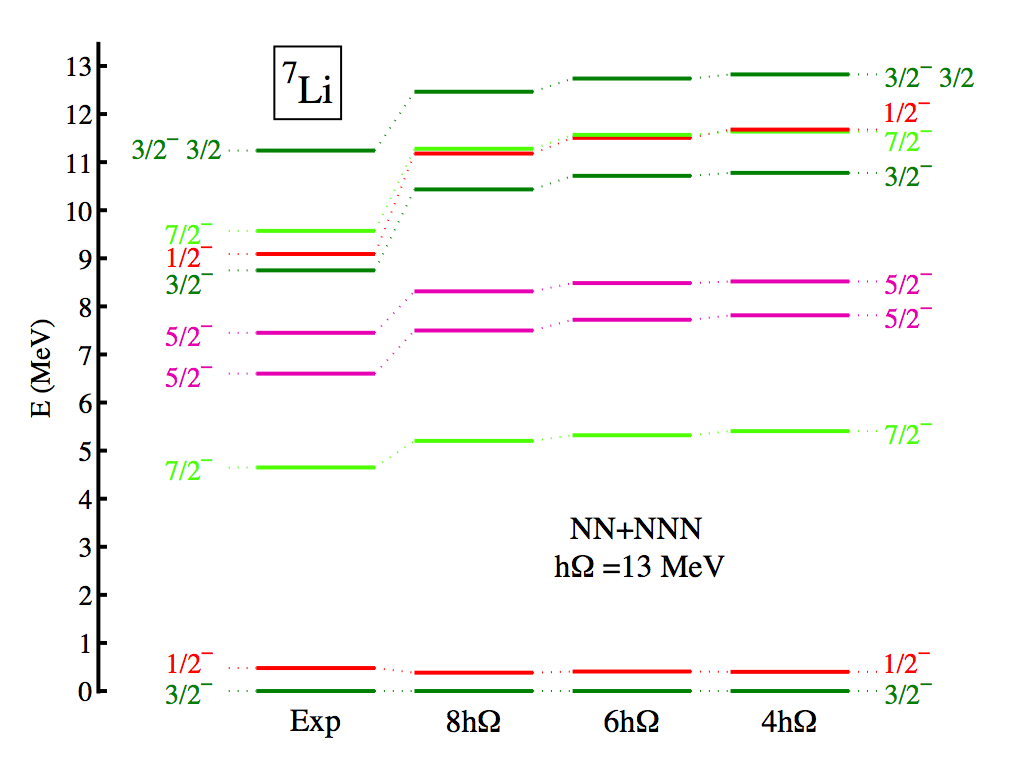}\includegraphics[width=0.5\textwidth]{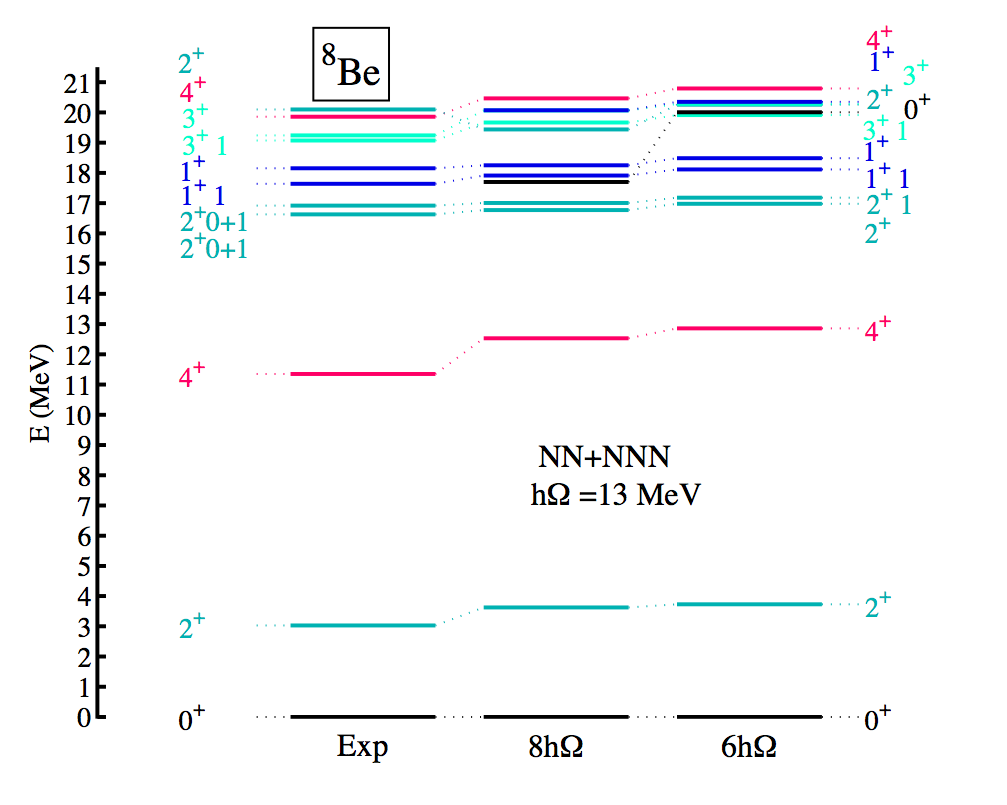}}
\caption{NCSM calculations of lowest-lying levels in 
$^7$Li and $^8$B using chiral  nuclear forces~\protect\cite{Maris:2012bt}.
[Image is reproduced with the permission of P. Maris.]}
\label{fig:ncsmA}      
\end{figure}

During the last several years, nuclear structure calculations have been
performed with  the chiral nuclear forces, leading to both postdictions and predictions
for nuclei to a given order in the expansion, and compared with
experiment, e.g. see Fig.~\ref{fig:ncsmA}.
The nuclear forces that are presently used in such calculations are constrained
by experimental measurements of NN scattering and light nuclei.
As the desired precision increases, which requires working to higher orders in the
expansion, the number of required experimental constraints increases.
Eventually, there are too few experimental constraints to practically reduce the systematic
uncertainty  below some level
in any given calculation.
However, LQCD calculations are expected to 
provide a way to constrain the nuclear forces
beyond what is possible with experiment, and hence to further reduce the
systematic uncertainties in nuclear structure calculations.
Beyond providing direct calculations of important quantities,
LQCD calculations of the light nuclei and nuclear forces can 
\begin{enumerate}
\item verify experimental constraints and/or reduce the uncertainties in the
  constraints imposed by experiment,
\item constrain components of the nuclear forces that are inaccessible to
  experiment, for instance the light-quark mass dependences which dictates some
  of the multi-pion
  vertices, and multi-neutron forces,
\item constrain counterterms at higher orders in the expansion to further
  reduce the systematic uncertainties.
\end{enumerate}


\section{Lattice QCD}

LQCD is a technique in which space-time is discretized into a
four-dimensional grid
and the QCD path integral over the quark and gluon fields at each
point in the grid is performed in Euclidean space-time
using Monte Carlo methods.
A LQCD calculation of a given quantity will deviate from its value in nature
because of the finite volume of the space-time (with $L^3\times T$ lattice points)
over which the fields exist, and
the finite separation between space-time points (the lattice spacing, $b$).
However, such deviations can be systematically removed by performing
calculations in multiple volumes with multiple lattice spacings, and
extrapolating using the
theoretically known functional dependences on each.
Supercomputers are needed for such calculations due to the number of space-time
points
and the Monte Carlo evaluation of the path integral over the dynamical fields.
In order for a controlled  continuum extrapolation, the lattice spacing  must be small enough
to resolve structures induced by the strong dynamics, 
encapsulated by $b\Lambda_\chi\ll 1$ where $\Lambda_\chi$ is the scale of
chiral symmetry breaking.
Further, in order to have the hadron masses, and also the scattering
observables, exponentially close to their infinite-volume values, the lattice
volume must be large enough to contain the lightest strongly
interacting particle, encapsulated by $m_\pi L \gsim 2\pi$ where $m_\pi$ is the
mass of the pion and $L$ is the extent of the spatial dimension of the cubic
lattice volume (this, of course, can be generalized to  non-cubic volumes).
Effective field theory (EFT) descriptions of these observables exist
for $b\Lambda_\chi\lsim  1$ (the Symanzik action and its translation into
chiral perturbation theory
($\chi$PT) and other frameworks) and $m_\pi L \gsim 2\pi$ (the p-regime of
$\chi$PT and other frameworks).
The low-energy constants in the appropriate EFT are fit to the results of the
LQCD calculations, which are then used to take the limit $b\rightarrow 0$ and
$L\rightarrow\infty$.
Computational resources devoted to LQCD calculations are becoming
sufficient to be able to perform calculations at the physical values of the
light quark masses in large enough volumes and at small enough lattice
spacings to be relevant, but the majority  of
present day calculations are performed with pion masses of $m_\pi\gsim 200~{\rm MeV}$.
Therefore, most calculations require the further extrapolation of
$m_q\rightarrow m_q^{\rm phys}$, but do not yet include strong isospin breaking or
electromagnetism.
In principle, the gauge-field configurations that are generated 
in LQCD calculations
can be used to
calculate an enormous array of observables, spanning the range from particle to
nuclear physics.  In practice, this is becoming less common, largely due to the
different scales relevant to particle physics and to nuclear physics.
Calculations of quantities involving the pion with a mass of 
$m_\pi\sim 140~{\rm  MeV}$ are substantially different from those of, say, the
triton with a mass of $M(^3{\rm H})\sim 3~{\rm GeV}$, and with the typical scale of
nuclear excitations being $\Delta E \sim 1~{\rm MeV}$.
Present day dynamical LQCD calculations of nuclear physics quantities are performed with
$m_\pi\sim 400~{\rm MeV}$, lattice spacings of $b\sim 0.1~{\rm fm}$ and volumes
with spatial extent of $L\sim 4~{\rm fm}$.

\begin{figure}[!ht]
\centerline{\includegraphics[width=0.6\textwidth]{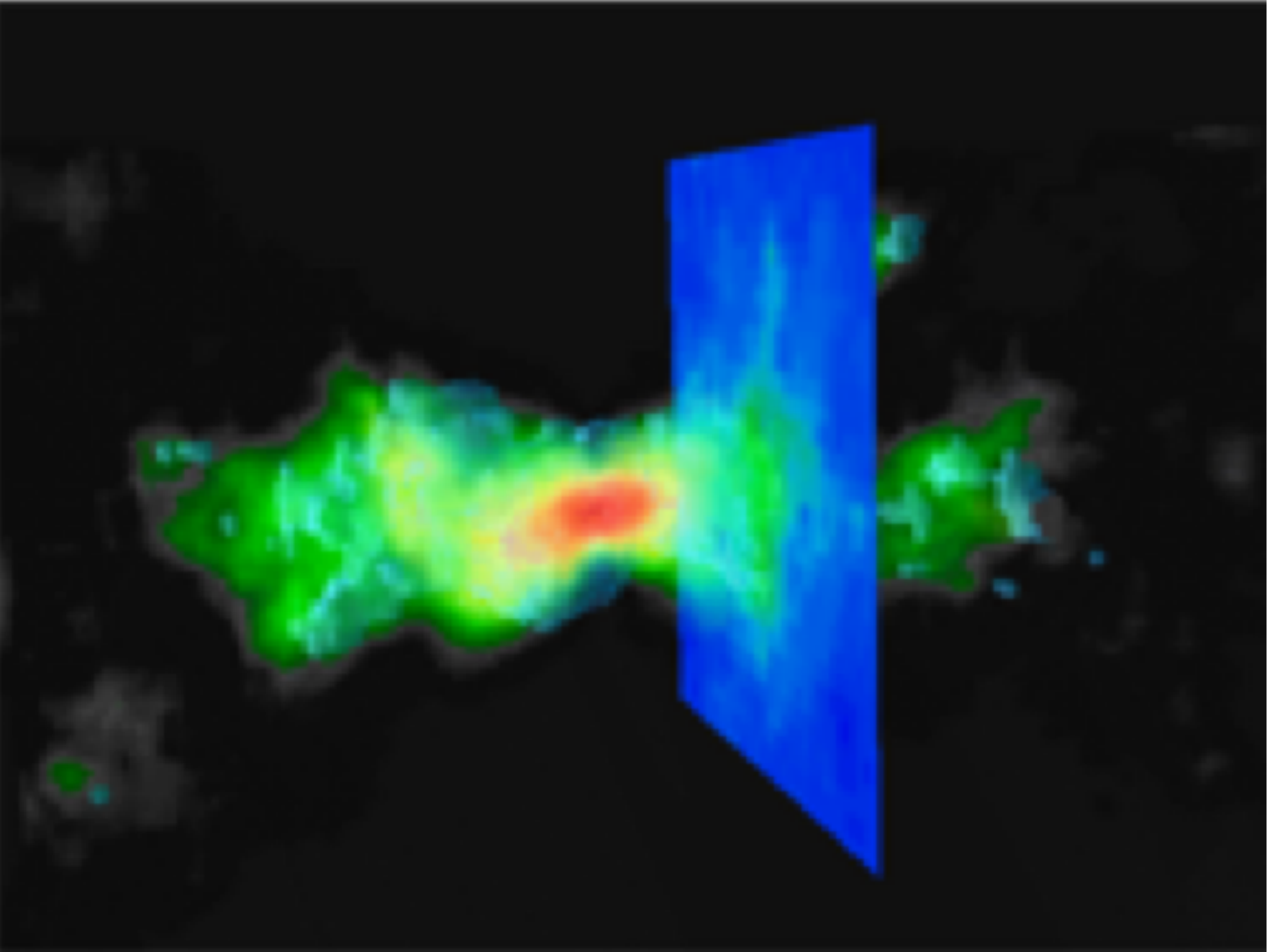}}
\caption{An example of (the real part of one component of)
a light-quark propagator.
The (blue) ``wall'' corresponds to the anti-periodic boundary conditions imposed in
the time direction.  [Image is reproduced with the permission of R. Gupta.]
\label{fig:prop}}
\end{figure}

LQCD calculations are approached in the same way that experimental efforts
use detectors to measure one or more quantities - the 
computer is equivalent to the accelerator
and the algorithms, software stack, and 
parameters of the LQCD calculation(s)
are the equivalent of  the detector.  The parameters, such as lattice spacing,
quark masses and volume, are selected based upon available computational
resources, and simulations of the precision of the calculation(s) required to
impact the physical quantity of interest, i.e. simulations  of the LQCD Monte
Carlos are performed.  The size of the computational resources required for
cutting edge calculations are such that you only get ``one shot at it''.
A typical work-flow of  a LQCD calculation consists of three major
components.  
The first component is the production of an ensemble of gauge-field
configurations which contain statistically independent samplings of the 
gluon fields resulting from the LQCD action.  
The production of gauge-fields requires the largest partitions on the
leadership class computational facilities, 
typically requiring  $\gsim 128$K compute cores.
Present-day calculations have
$n_f=0, 2, 2+1, 3, 2+1+1$ dynamical light-quark flavors and use the Wilson, ${\cal
  O}(b)$-improved-Wilson, staggered (Kogut-Susskind), domain-wall or overlap
discretizations, each of which have their own ``features''.  
It is the
evaluation of the light-quark determinant 
(the determinant of a sparse matrix with dimensions
$\gsim 10^8\times 10^8$)
that consumes the largest fraction
of the resources.
Roughly speaking, $\gsim 10^4$ Hybrid Monte Carlo (HMC) trajectories are required to produce
an ensemble of $10^3$ decorrelated gauge fields, but in many instances this is
an under estimate.
For observables involving
quarks, a second component 
of production
is the determination of the light-quark
propagators on each of the configurations. 
The light-quark
propagator from a  given source point 
(an example of which is shown in Fig.~\ref{fig:prop})
is determined
by an iterative inversion of the quark two-point function, 
using the conjugate-gradient (CG) algorithm or variants thereof such as
BiCGSTAB,
or the most recently developed multi-grid (MG).
During the last couple of years, the propagator production codes have been
ported to run on GPU machines in parallel.  
GPU's can perform propagator
calculations faster than standard CPU's by an order of magnitude, and
have led to a major reduction in the statistical uncertainties in many calculations.
There have been numerous algorithm developments that have also reduced the
resources required for propagator production, such as the implementation of
deflation techniques and the use of multi-grid methods.
The third component of a LQCD calculation is the production of correlation
functions from the light-quark propagators.  This involves performing all of
the Wick contractions that contribute to a given quantity.  
The number of contractions required for computing a
single hadron correlation function is small.
However,  to acquire long plateaus
in the effective mass plots (EMPs) that
persist to short times, L\"uscher-Wolff type 
methods involve the
computation of a large number of correlation functions resulting from different
interpolating operators, and the number of contractions
can  become large.
In contrast, the naive number of contractions required for  a nucleus quickly
becomes astronomically large ($\sim 10^{1500}$ for uranium), but symmetries in
the contractions, and new algorithms (e.g. Ref.~\cite{Detmold:2012eu}) greatly reduce the
number of operations that must be performed.  
A further consequence of the hierarchy of mass scales is that there is an
asymptotic signal-to-noise problem in nuclear correlation functions.  
The ratio of the mean value of the correlation function to the variance of the
sample from which the mean is evaluated degrades exponentially at large times.
However, this is absent at short and intermediate times and
the exponential degradation of the signal-to-noise in the 
correlation functions can be avoided.

\section{Cold Nuclear Physics with Lattice QCD}

Capability computing resources provided by leadership class computing
facilities are used to produce ensembles of gauge-field configurations,
while capacity computing resources, both those operated by USQCD and elsewhere
are used to perform observable-dependent calculations of correlation functions
using these
configurations.
Thus the capability resources enable a multitude of physics calculations to be
accomplished with the capacity resources.
In the area of cold nuclear physics there is currently a  well-defined set of
goals, and a program in place to accomplish  these goals, as described in one
of the 2013
USQCD Whitepapers~\cite{USQCDwp}: 
{\bf Hadron Structure}, 
{\bf Hadron Spectroscopy},
{\bf Hadronic Interactions, Nuclear Forces and Nuclei},
and 
{\bf Fundamental Symmetries}.

\subsection{The Spectra and Structure of the Hadrons}

\begin{figure}[!ht]
\centerline{\includegraphics[width=0.7\textwidth]{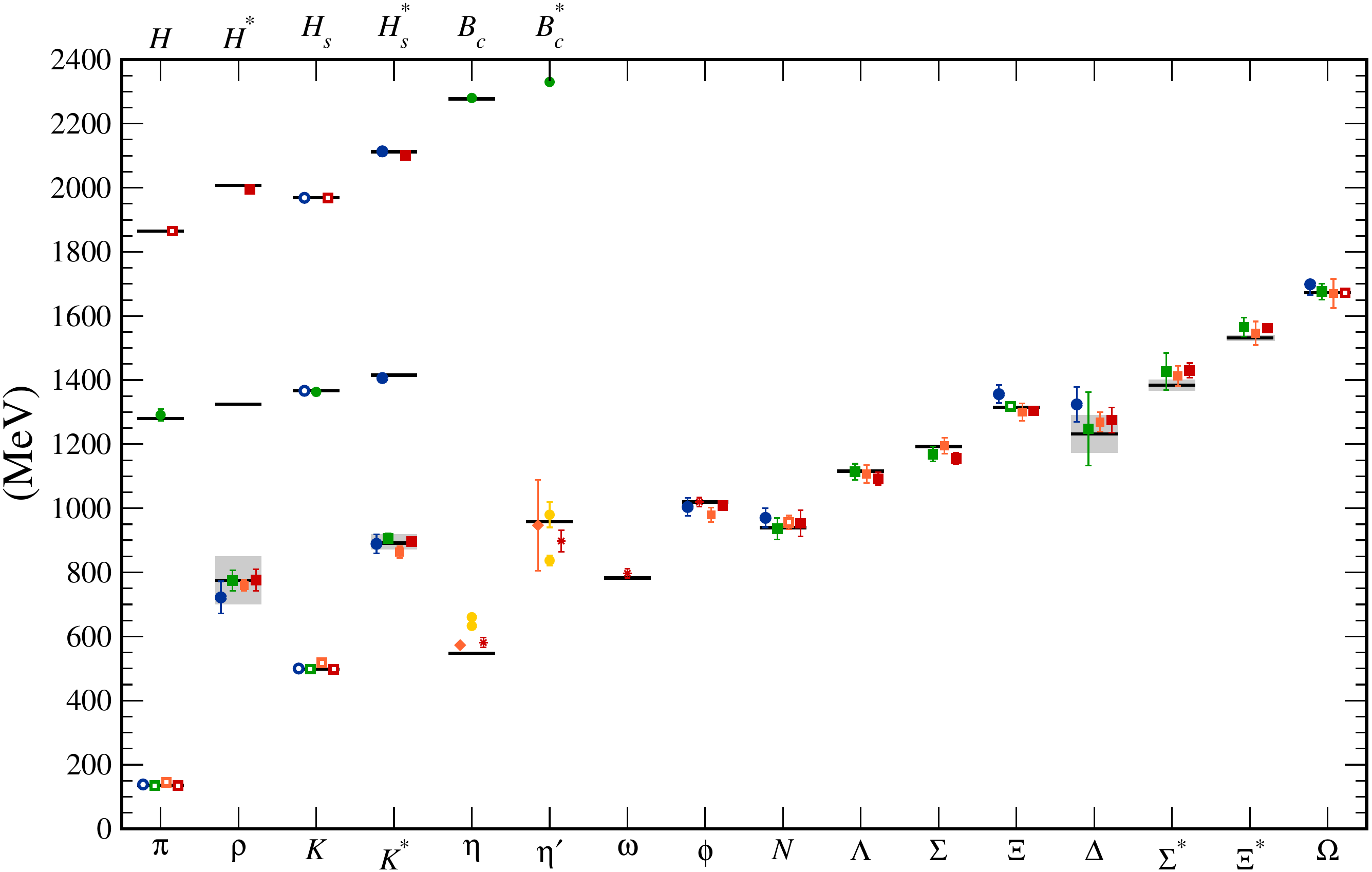}}
\caption{A summary of the low-lying hadron masses calculated with LQCD~\protect\cite{Kronfeld:2012uk,Kronfeld:2012ym}.  
[Image is reproduced with the permission of A. Kronfeld.]
\label{fig:spectrum}}
\end{figure}

\begin{figure}[!ht]
\centerline{\includegraphics[width=0.7\textwidth]{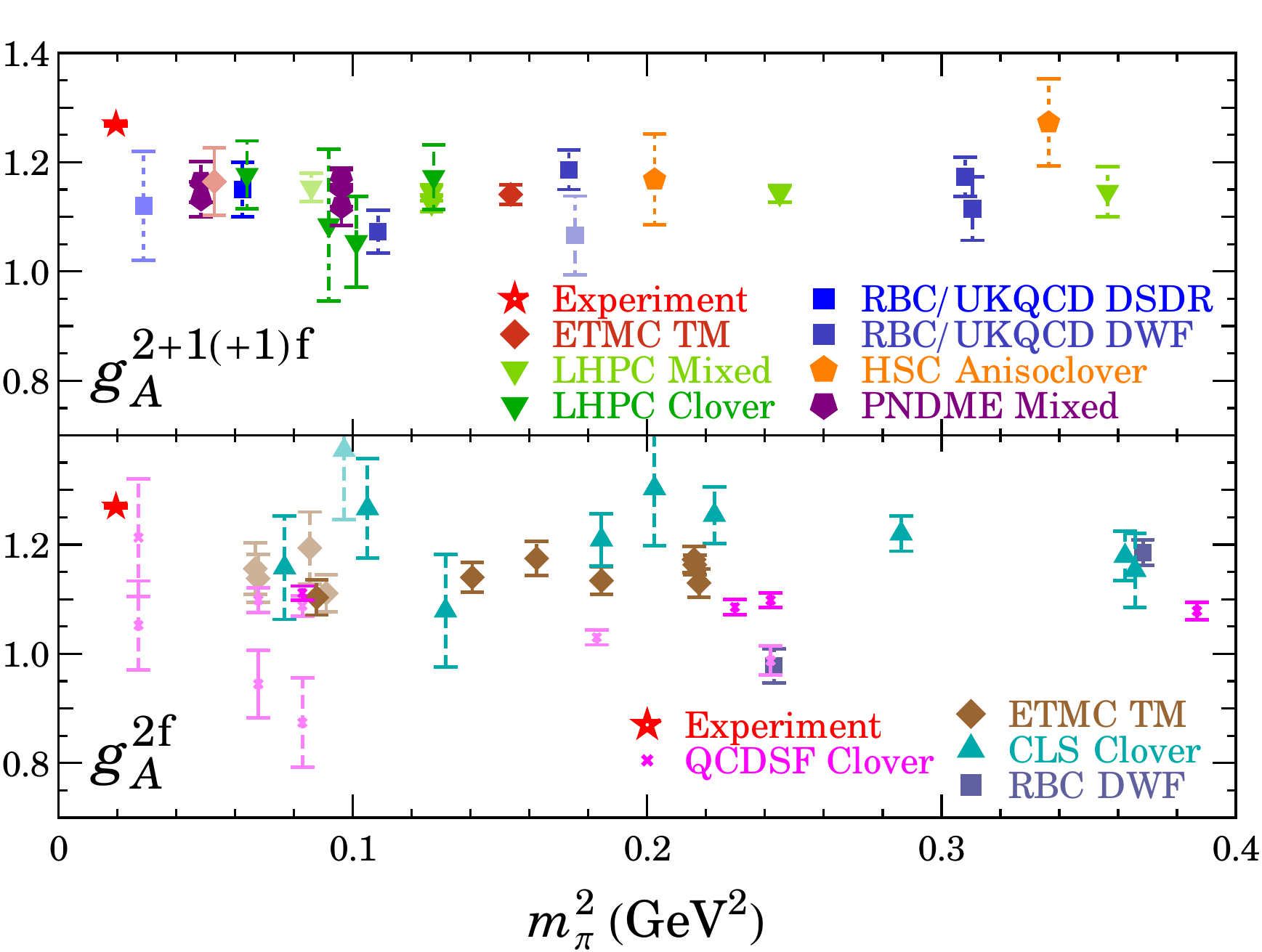}}
\caption{A summary of LQCD calculations of $g_A$.~\protect\cite{Lin:2012ev}. 
[Image is reproduced with the permission of H.-W. Lin.]
\label{fig:gA}
}
\end{figure}

Before calculations of nuclei can be sensibly undertaken, the mass and
structure of the nucleon
must be reproduced in LQCD calculations.
The spectrum of the lowest-lying hadrons calculated with LQCD 
is shown in Fig.~\ref{fig:spectrum},
from which we observe that indeed LQCD is postdicting all of the light-hadron
masses within uncertainties.
Beyond its mass, 
one property of the nucleon that is well known experimentally is the forward-matrix
element of the isovector axial current, $g_A$.
Significant effort has been put into calculating $g_A$ with LQCD, a summary of
which is shown in Fig.~\ref{fig:gA}, but the extrapolated LQCD value has consistently been 
smaller than the experimental value.
With calculations beginning to be  performed
at the physical pion mass, the community is focused on understanding and
quantifying the systematic uncertainties in these calculations.

\begin{figure}[!ht]
\centerline{\includegraphics[width=0.85\textwidth]{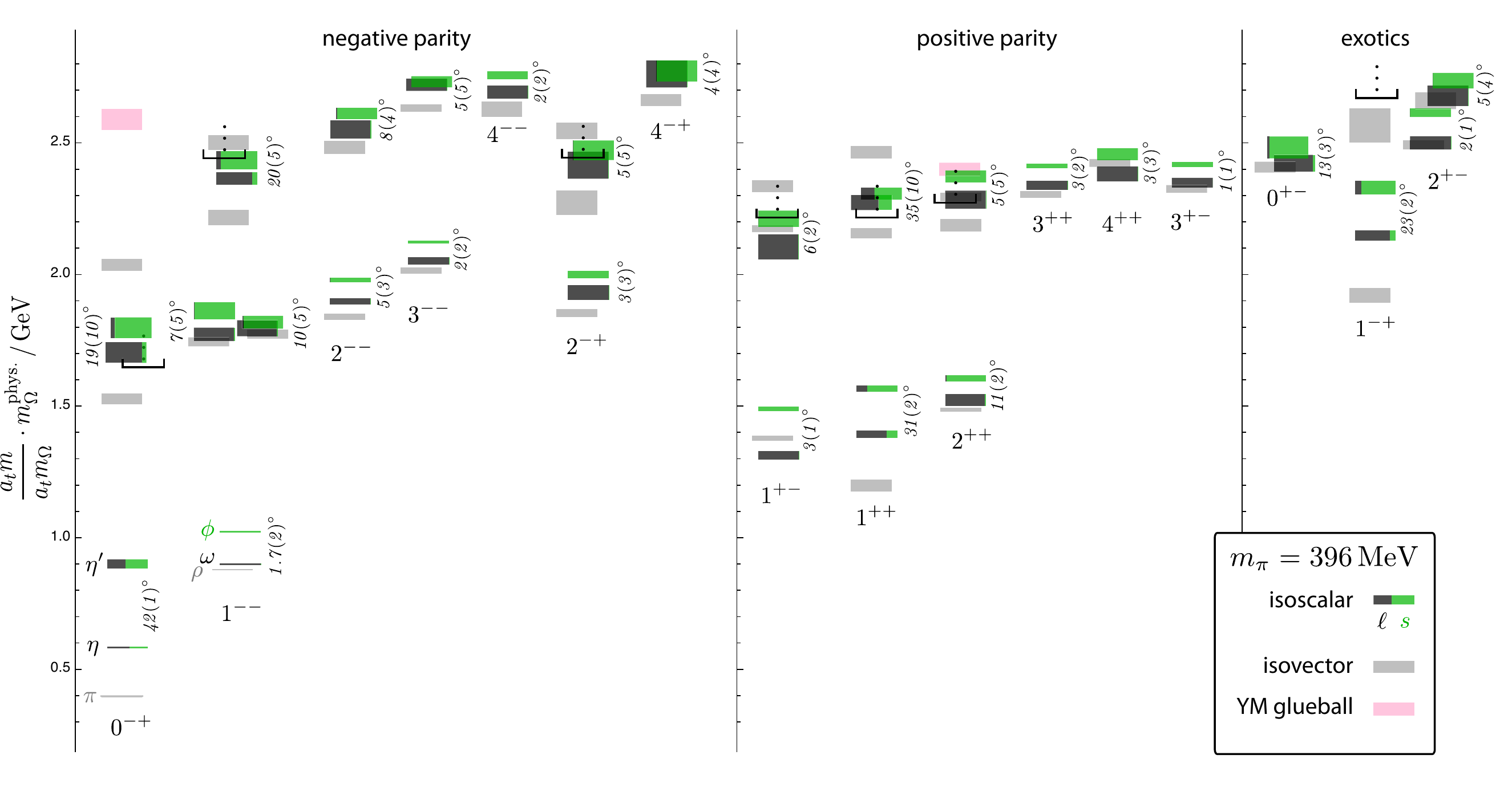}}
\caption{The spectra of isoscalar mesons calculated at $m_\pi\sim 396~{\rm
    MeV}$ by the JLab LQCD group~\protect\cite{Dudek:2011tt}
[Image is reproduced with the permission of R. Edwards.]
\label{fig:Isoscalars}
}
\end{figure}

A central element of the physics program at JLab is
to determine  the excited spectra of mesons and baryons, including
searching for exotic states that are beyond the naive nonrelativistic quark
model of hadrons, but arise naturally in QCD. A critical component of this
program is the LQCD calculations of the spectra.
They will play a central role
in interpreting and understanding the experimental measurements.  The spectra
of such states is complicated by the presence of open multi-hadron channels and
significant formal developments remain to be put in place before rigorous
statements about the spectra can be made.  Calculations at unphysical pion
masses have been performed by the JLab LQCD group, examples of which are
shown in Fig.~\ref{fig:Isoscalars}, and remarkable progress has been made in
the identification of states in these calculations.
The aim is to have LQCD predict the exotic spectra of hadrons before, or
at the same time as, the GlueX experiment at JLab
runs, targeting the 2018 milestone HP15.

\subsection{Meson-Meson Scattering}

\noindent 
Multi-hadron LQCD calculations are significantly more challenging than 
single-hadron calculations for a number of reasons, and systems involving baryons
are even more challenging.
Meson-meson systems are the simplest 
multi-hadron systems, and impressive progress has been made in the recent past, particularly
when the LQCD calculations are combined with
$\chi$PT.
There is little or no signal-to-noise problem 
in such calculations and
therefore highly accurate LQCD calculations 
of stretched-isospin states
can be performed
with modest computational resources. Moreover, the EFTs which describe the
low-energy interactions of pions and kaons, including lattice-spacing
and finite-volume effects, have been developed to non-trivial orders
in the chiral expansion. 
\begin{figure}[!ht]
\centerline{\includegraphics[width=0.56\textwidth]{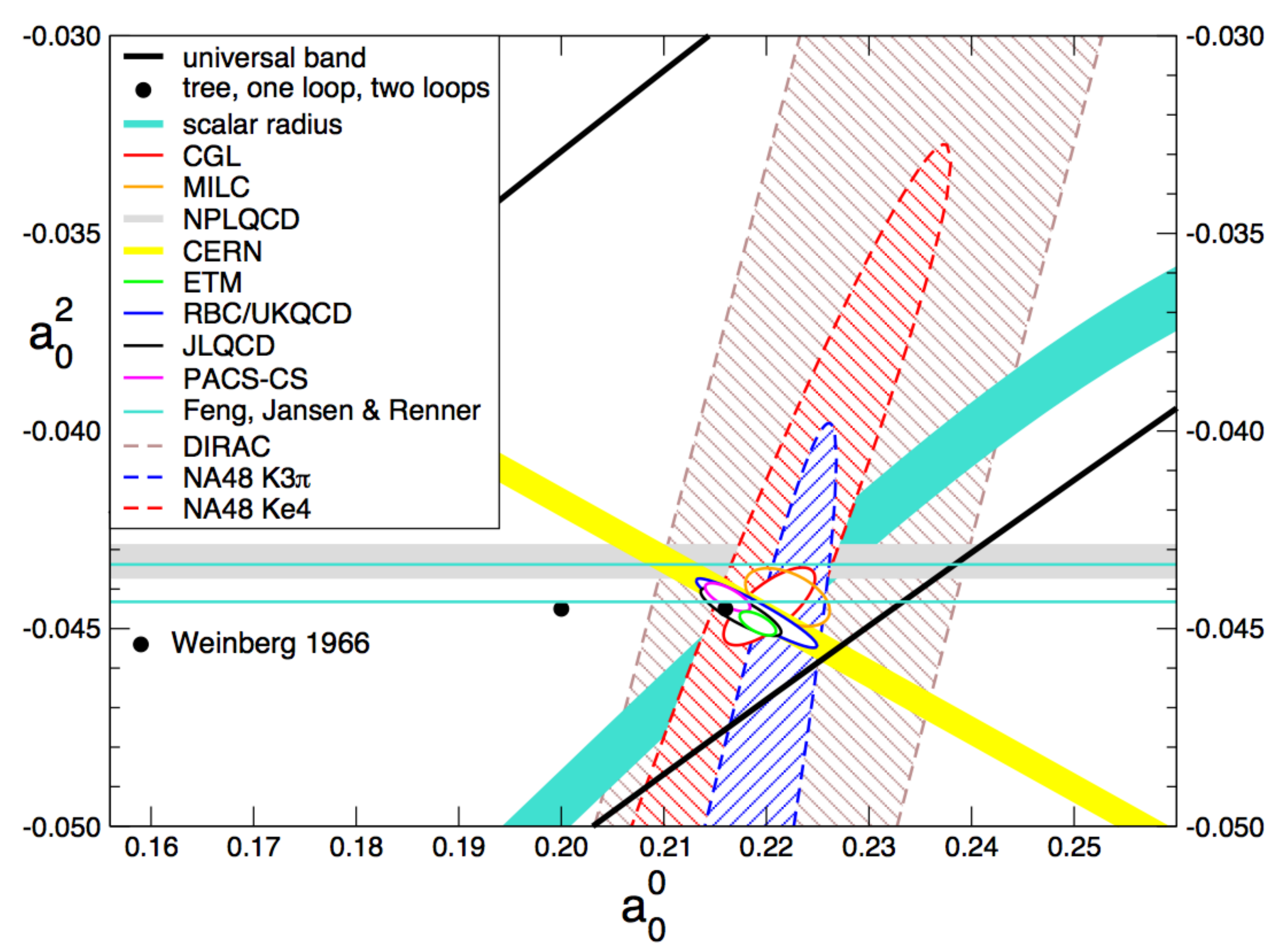}
\ \ \includegraphics[width=0.34\textwidth]{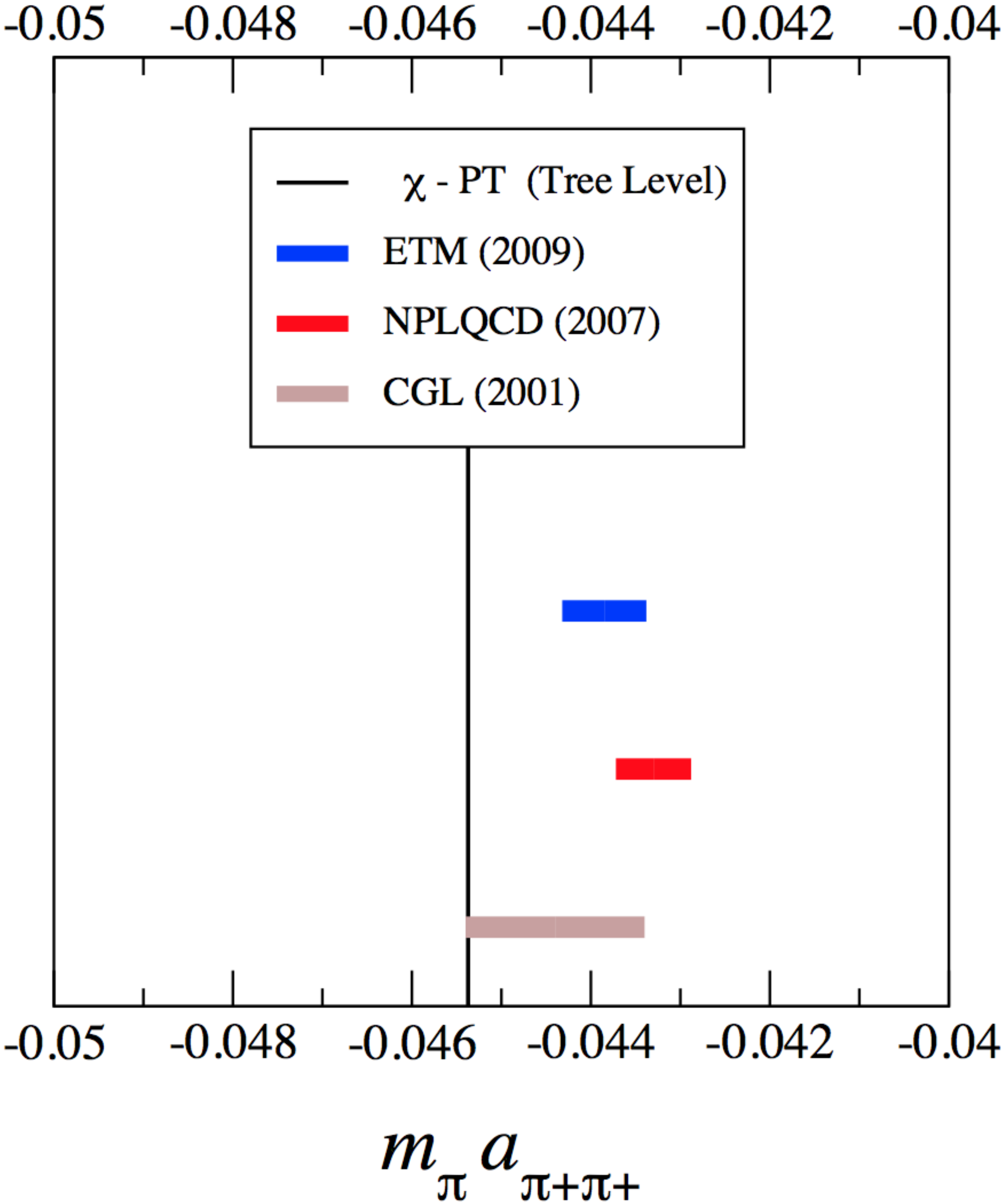}}
\caption{Constraints on threshold s-wave $\pi\pi$ scattering ~\protect\cite{Leutwyler:2008fi}. 
[Image in the left panel is reproduced with the permission of H. Leutwyler.]
\label{fig:pipi2}
}
\end{figure}
The $I=2$ pion-pion ($\pi^+\pi^+$) scattering length serves as a
benchmark calculation with an accuracy that can only be aspired to
in other systems.   The scattering
lengths for $\pi\pi$ scattering in the s-wave are uniquely predicted
at LO in $\chi$PT~\cite{Weinberg:1966kf}:
\begin{eqnarray}
m_{\pi^+} a_{\pi\pi}^{I=0} \ = \ 0.1588 \ \ , 
\ \ m_{\pi^+} a_{\pi\pi}^{I=2} \ = \
-0.04537 
\ \ \ .
\label{eq:CA}
\end{eqnarray}
While experiments 
do not directly provide stringent
constraints on the scattering lengths, a determination of s-wave
$\pi\pi$ scattering lengths using the Roy equations has reached a
remarkable level of
precision~\cite{Colangelo:2001df,Leutwyler:2008fi}:
\begin{eqnarray}
m_{\pi^+} a_{\pi\pi}^{I=0} \ = \ 0.220\pm 0.005 \ \ , 
\ \ m_{\pi^+} a_{\pi\pi}^{I=2} \ = \ -0.0444\pm 0.0010
\ \ \ .
\label{eq:roy}
\end{eqnarray}
The Roy equations~\cite{Roy:1971tc} use dispersion theory to relate
scattering data at high energies to the scattering amplitude near
threshold. 
At present, LQCD can compute $\pi\pi$ scattering
only in the $I=2$ channel with precision 
as the $I=0$ channel contains disconnected
diagrams which require large computational resources. 
It is of great interest to compare the precise Roy
equation predictions with LQCD
calculations,
and  Fig.~\ref{fig:pipi2} summarizes theoretical and
experimental constraints on the s-wave $\pi\pi$ scattering
lengths~\cite{Leutwyler:2008fi}. 
This is clearly a strong-interaction
process for which  theory has somewhat out-paced the challenging experimental
measurements.

Mixed-action $n_f=2+1$ LQCD calculations, 
employing domain-wall valence quarks on a rooted staggered sea and 
combined with mixed-action $\chi$PT,
have predicted~\cite{Beane:2007xs}
\begin{eqnarray}
m_{\pi^+} a_{\pi\pi}^{I=2} & = &  -0.04330 \pm 0.00042
\ \ \ ,
\label{eq:nplqcd2}
\end{eqnarray}
at the physical pion mass.
The agreement between this result and the Roy equation
determination is a striking confirmation of the lattice methodology,
and a powerful demonstration of the constraining power of chiral
symmetry in the meson sector. 
However, LQCD calculations at one or more smaller lattice spacings, 
and with different discretizations,
are
required to verify and further refine this calculation.
The ETM collaboration has  performed a $n_f=2$ calculation of
the $I=2$ $\pi\pi$ scattering length~\cite{Feng:2009ij}, producing a 
result extrapolated to the physical pion mass of
\begin{eqnarray}
m_{\pi^+} a_{\pi\pi}^{I=2} & = &  -0.04385 \pm 0.00028 \pm 0.00038
\ \ \ .
\label{eq:TMpipi}
\end{eqnarray}

\begin{figure}[!ht]
\centerline{\includegraphics[width=0.45\textwidth]{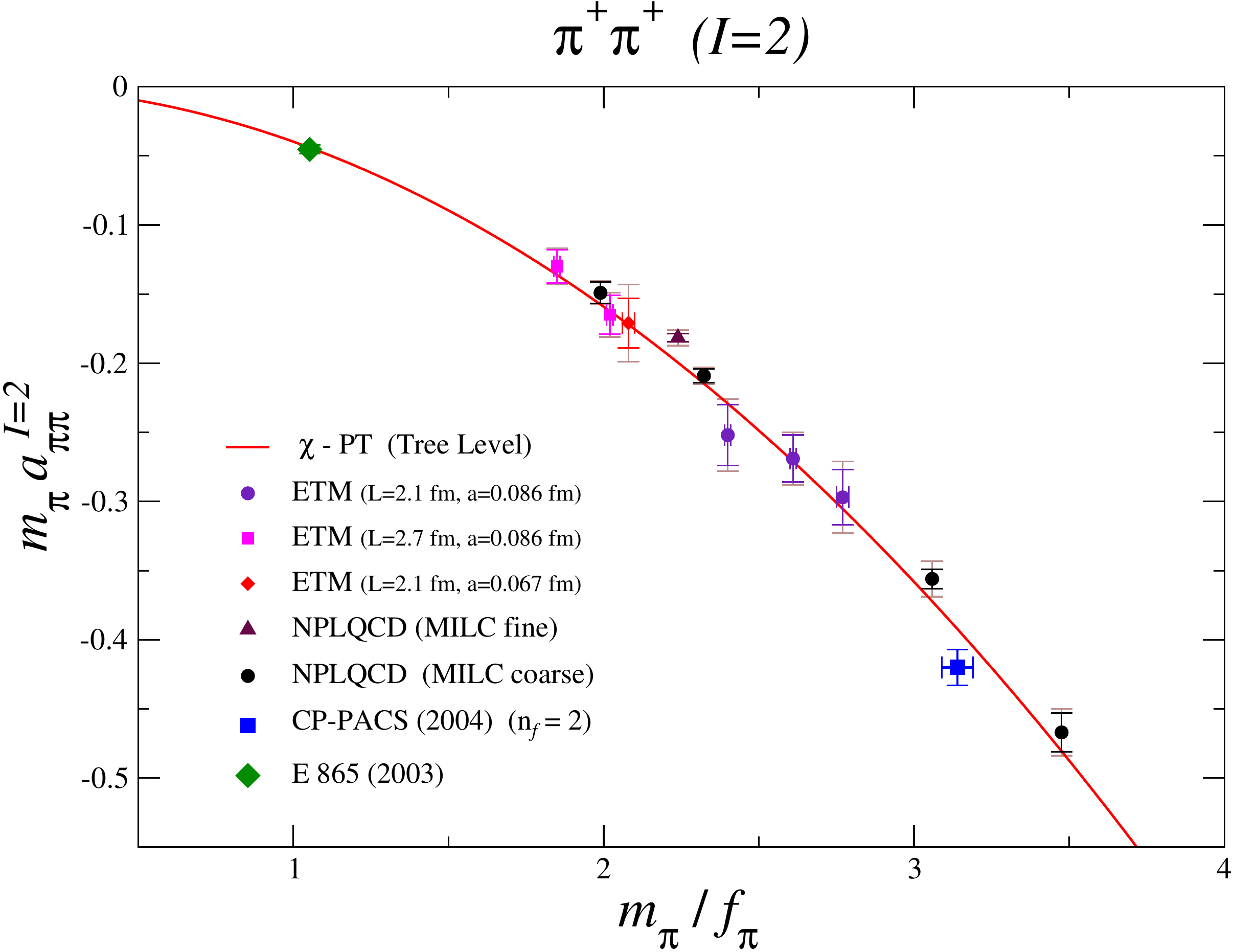}
\ \ \includegraphics[width=0.45\textwidth]{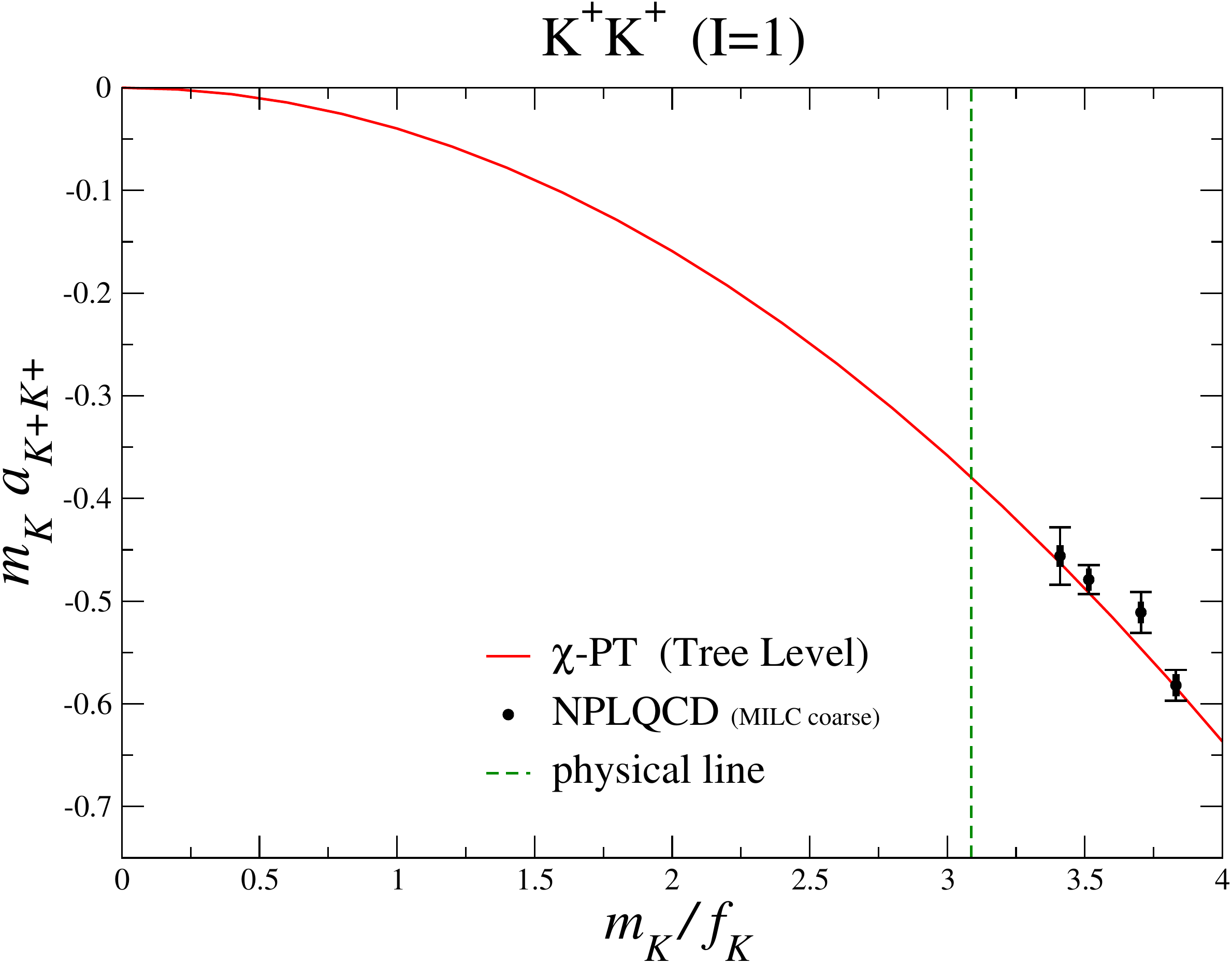}}
\caption{$m_{\pi^+} a_{\pi^+\pi^+}$ vs. $m_{\pi^+}/f_{\pi^+}$ (left panel)
and $m_{K^+} a_{K^+K^+}$ vs. $m_{K^+}/f_{K^+}$ (right panel). 
The solid (red) curves are the current algebra predictions.
\label{fig:CAplots}
}
\end{figure}

It is interesting to compare the pion mass dependence of the
meson-meson scattering lengths to the current algebra
predictions.  In Fig.~\ref{fig:CAplots} (left panel) one sees that the
$I=2$ $\pi\pi$ scattering length is consistent with 
the current algebra result up
to pion masses that are expected to be at the edge of the chiral
regime in the two-flavor sector. While in the two-flavor theory one
expects fairly good convergence of the chiral expansion and, moreover,
one expects that the effective expansion parameter is small in the
channel with maximal isospin, the LQCD calculations clearly imply a
degree of 
cancellation between chiral logs and counterterms. 
However, as one sees in Fig.~\ref{fig:CAplots} (right
panel), the same phenomenon occurs in $K^+K^+$ where the chiral
expansion is governed by the strange quark mass and is therefore
expected to be much more slowly converging.   This remarkable conspiracy between
chiral logs and counterterms for the meson-meson scattering lengths remains
mysterious.

\begin{figure}[!ht]
\centerline{\includegraphics[width=0.45\textwidth]{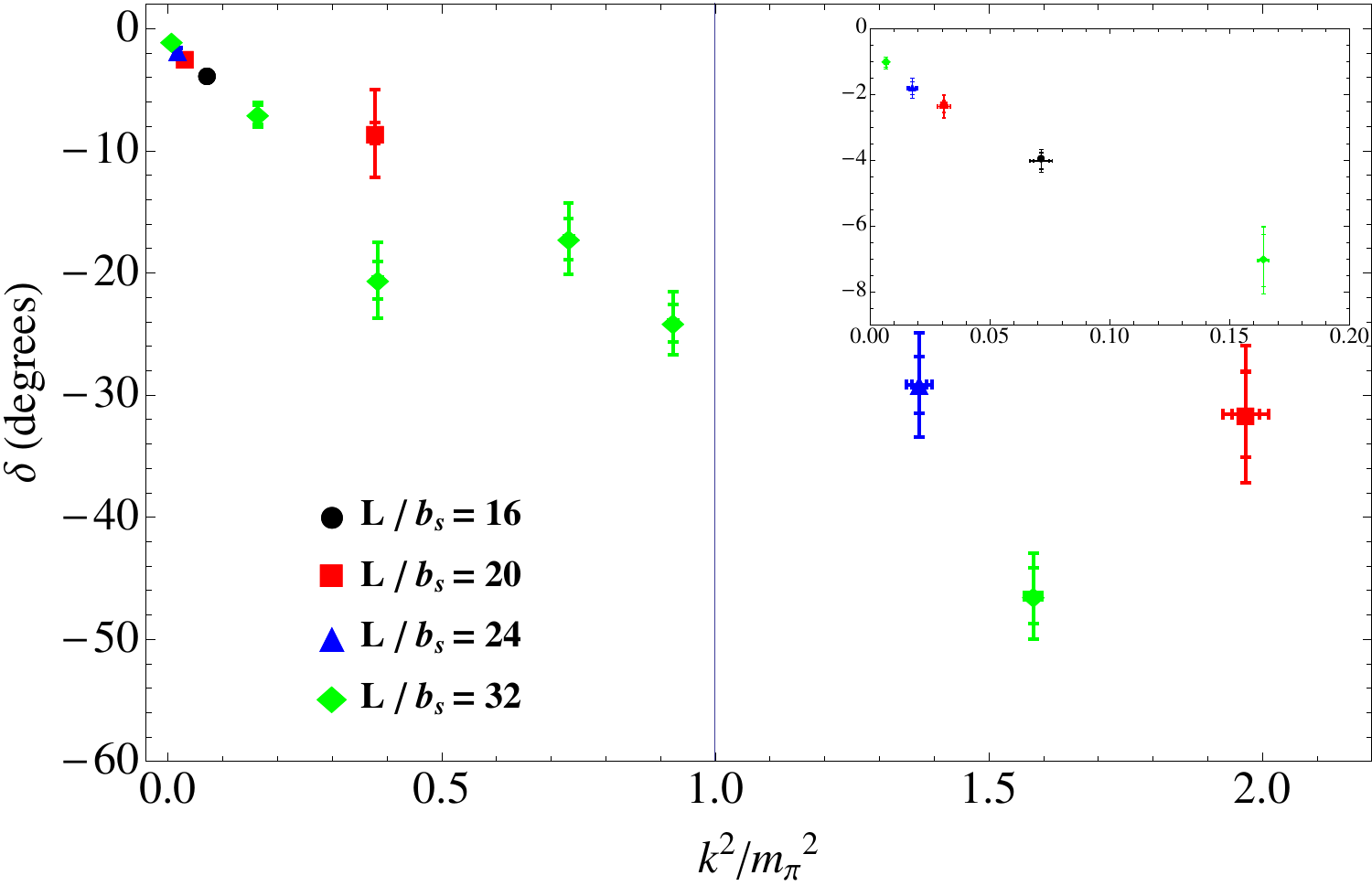}
\ \ \includegraphics[width=0.45\textwidth]{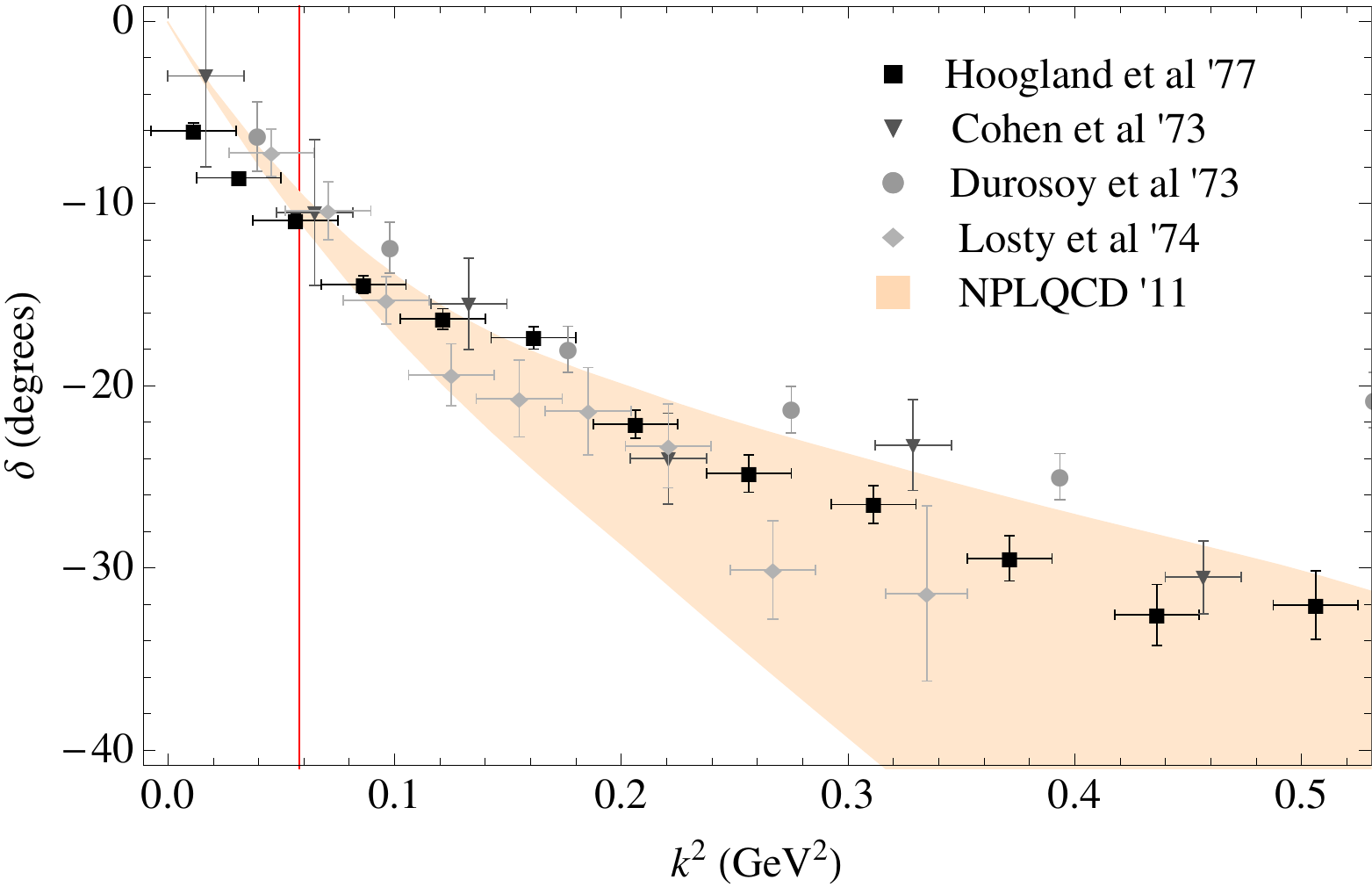}}
\caption{The $\pi^+\pi^+$ scattering phase-shift.  
The left panel shows the results of
the LQCD calculations below the inelastic threshold 
($|{\bf k}|^2 = 3 m_\pi^2$)
at a pion mass of
$m_\pi\sim 390~{\rm MeV}$~\protect\cite{Beane:2011sc}.  
The vertical (blue) line denotes the
start of the t-channel cut.
The shaded region in the 
right panel shows the results of the LQCD calculation extrapolated to the
physical pion mass using NLO $\chi$PT, while the points and uncertainties
corresponds to the existing experimental data.
The vertical (red) line corresponds to the inelastic threshold.
\label{fig:pipidelta}
}
\end{figure}

\begin{figure}[!ht]
\centerline{\includegraphics[width=0.7\textwidth]{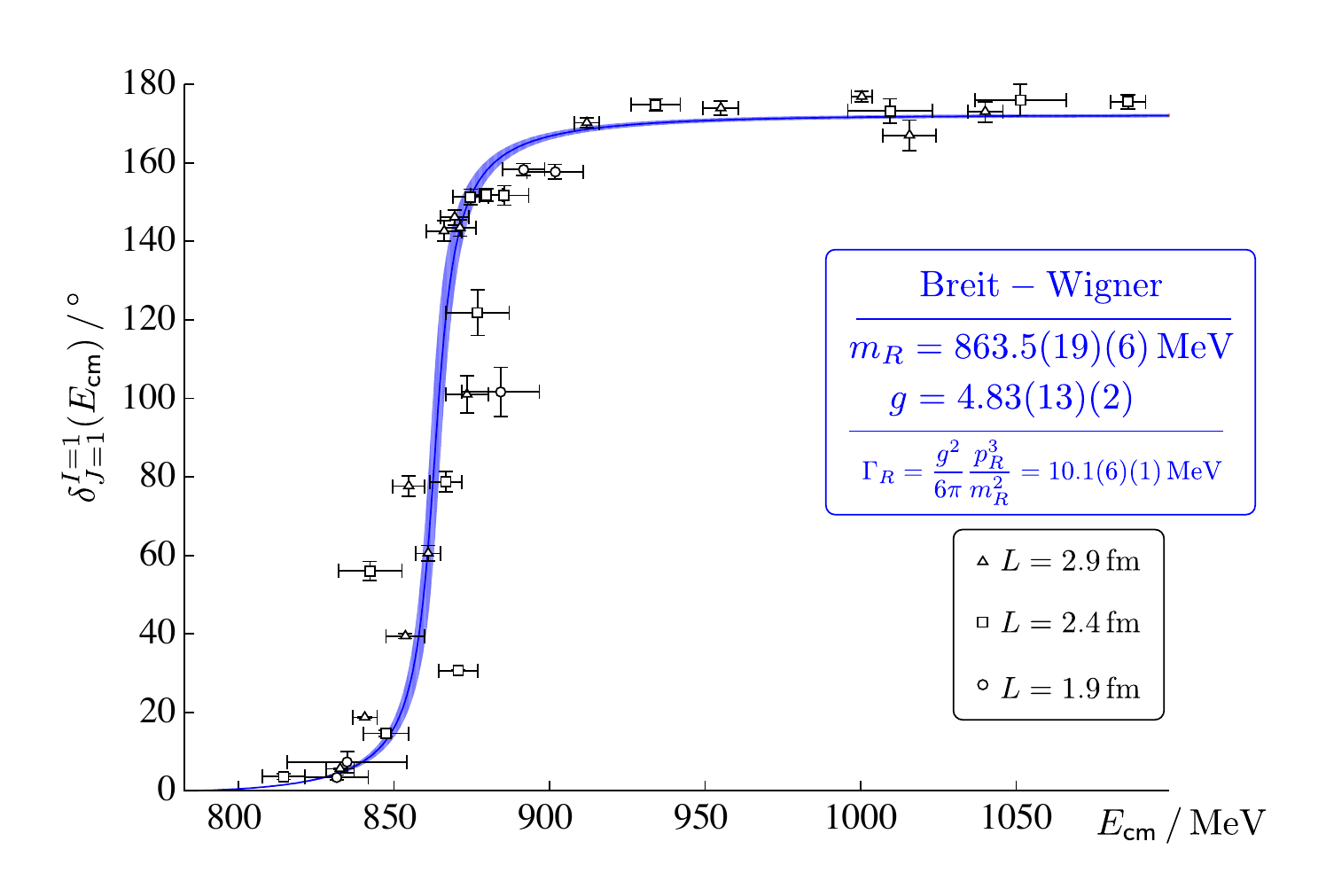}
}
\caption{The $\rho$-resonance at a pion mass of 
$m_\pi\sim 390~{\rm MeV}$~\protect\cite{Dudek:2012xn}
[Image is reproduced with the permission of R. Edwards.]
\label{fig:rho}
}
\end{figure}

LQCD calculations of the meson-meson scattering phase-shifts are much less
advanced than of the scattering length.
This is because the calculation of the phase shift, $\delta(E)$, at a
given energy, $E$, requires a LQCD calculation of the two-meson correlation
function at the energy $E$.
Generally speaking, a given calculation can determine the lowest few two-hadron
energy eigenvalues for a given momentum of the center-of-mass, and  that  multiple lattice volumes
will  allow for additional  values of $E$ at which to determine $\delta(E)$.
The first serious calculation of the s-wave ($l=0$) $I=2$ $\pi\pi$ phase-shift was done by the
CP-PACS collaboration with $n_f=2$ at a relatively large pion
mass~\cite{Yamazaki:2004qb}, 
and more 
recently two groups have performed calculations at lower pion 
masses~\cite{Dudek:2010ew,Beane:2011sc}, 
the results of which are shown in Fig.~\ref{fig:pipidelta}.
Further, in some nice work by the Hadron Spectrum Collaboration (HSC), 
the first efforts have been made to extract 
the d-wave ($l=2$)  $I=2$ $\pi\pi$ phase shift~\cite{Dudek:2010ew}.
One of the more exciting recent results is the mapping out of the
$\rho$-resonance 
at $m_\pi\sim 390~{\rm MeV}$
from the $\pi^+\pi^0$ energy-levels using Luscher's method, as
shown in Fig.~\ref{fig:rho}~\cite{Dudek:2012xn}.
%

\subsection{Nuclear Interactions}

\begin{figure}[!ht]
\centerline{\includegraphics[width=0.5\textwidth]{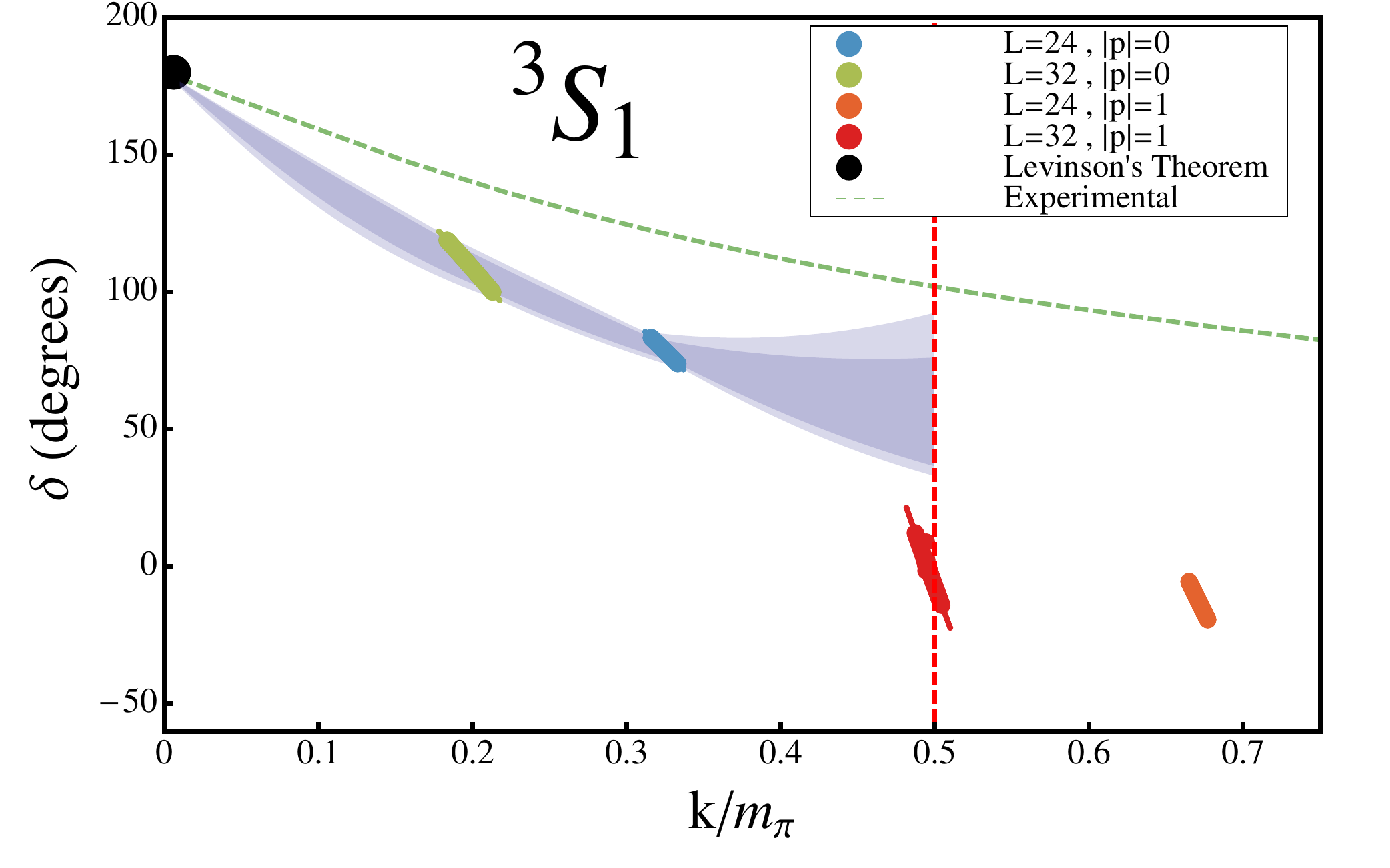}\ \
  \includegraphics[width=0.5\textwidth]{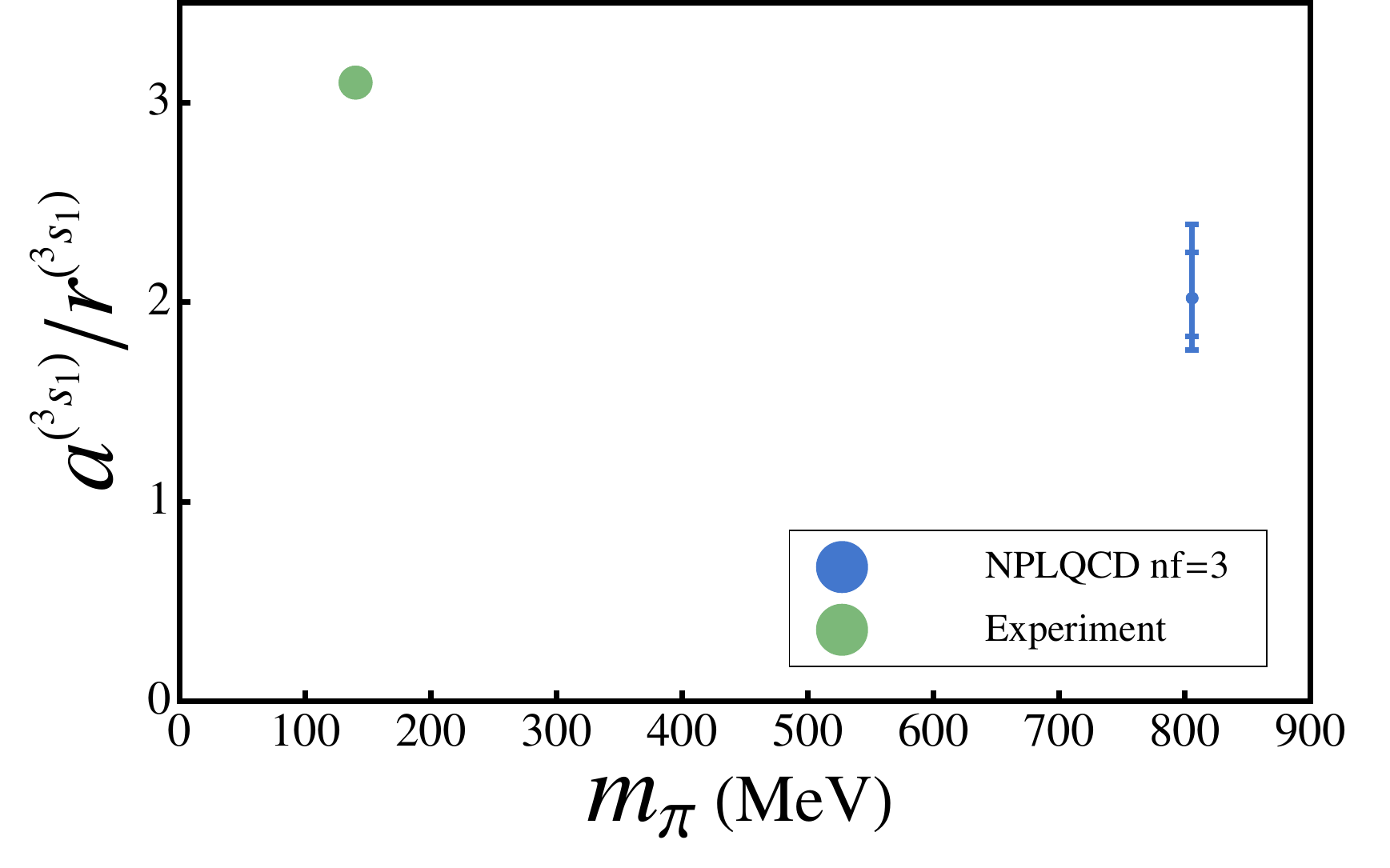}
}
\caption{The left panel shows the NN scattering phase shift in the $\siii$
  channel extracted from LQCD calculations at the SU(3) symmetric point,
  including the fit to the ERE at N$^2$LO.  
The right panel shows the ratio of the scattering length to effective range, a
quantity that is a measure of the naturalness of the system.
\label{fig:NNscattNPLQCD}
}
\end{figure}

Calculations of the nucleon-nucleon scattering lengths have been successfully underway for
the last decade~\cite{Fukugita:1994na,Fukugita:1994ve,Beane:2006mx,Beane:2009py,Ishii:2006ec,Aoki:2008hh,Aoki:2009ji,
  Yamazaki:2011nd,Yamazaki:2009ua,deForcrand:2009dh,Beane:2011iw,Inoue:2011ai,Beane:2012vq,Yamazaki:2012hi} 
for a range of pion masses.
Recently, LQCD calculations have been performed 
at  $m_\pi\sim 800~{\rm MeV}$
that also provide
the effective ranges~\cite{Beane:2013br}, the
results of which are shown in Fig.~\ref{fig:NNscattNPLQCD}.
Also shown are fits to the effective range expansion (ERE), including the shape parameter.
The scattering length and effective range in the $\siii$ channel
determined from the NLO fit to the ERE are
\begin{eqnarray}
m_\pi a^{(\siii)} & = & \ampitrip
\ \ \ ,\ \ \  
m_\pi r^{(\siii)}  \ = \ \rmpitrip
\ \ \ ,
\nonumber\\
a^{(\siii)}  & = & \ampitripphys~{\rm fm}
\ \ \ , \ \ \ \
r^{(\siii)}  \ = \ \rmpitripphys~{\rm fm}
\ \ \ .
\end{eqnarray}
The shape parameter obtained from the NNLO fit to the ERE expansion is:
$P m_\pi^3 = 2^{+5}_{-6}{}^{+5}_{-6}$.
An interesting aspect of this result is that the ratio of scattering length to effective
range, a measure of the naturalness of the system, is $\sim 2$, which is to be
compared with $\sim 3$ at the physical quark masses.  This leads one to
speculate that the deuteron might be unnatural over a large range of quark
masses and not just close to the physical values, indicating that it is not
finely tuned.  This speculation requires precise calculations at lighter quark
masses to determine if this is, in fact, the situation.

\subsection{Nuclei}

\begin{figure}[!ht]
  \centering
     \includegraphics[width=0.32\textwidth]{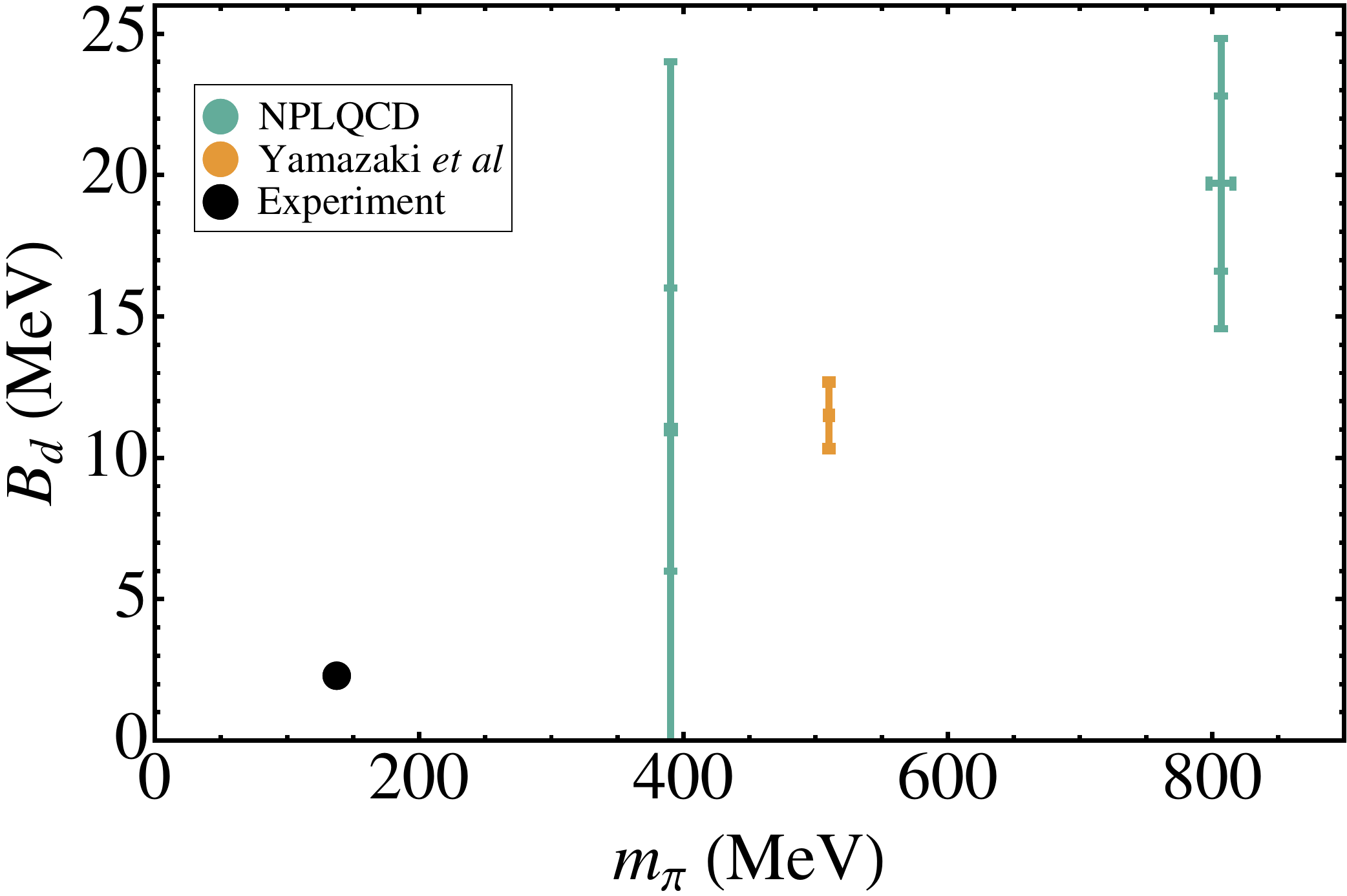}\ \ 
     \includegraphics[width=0.32\textwidth]{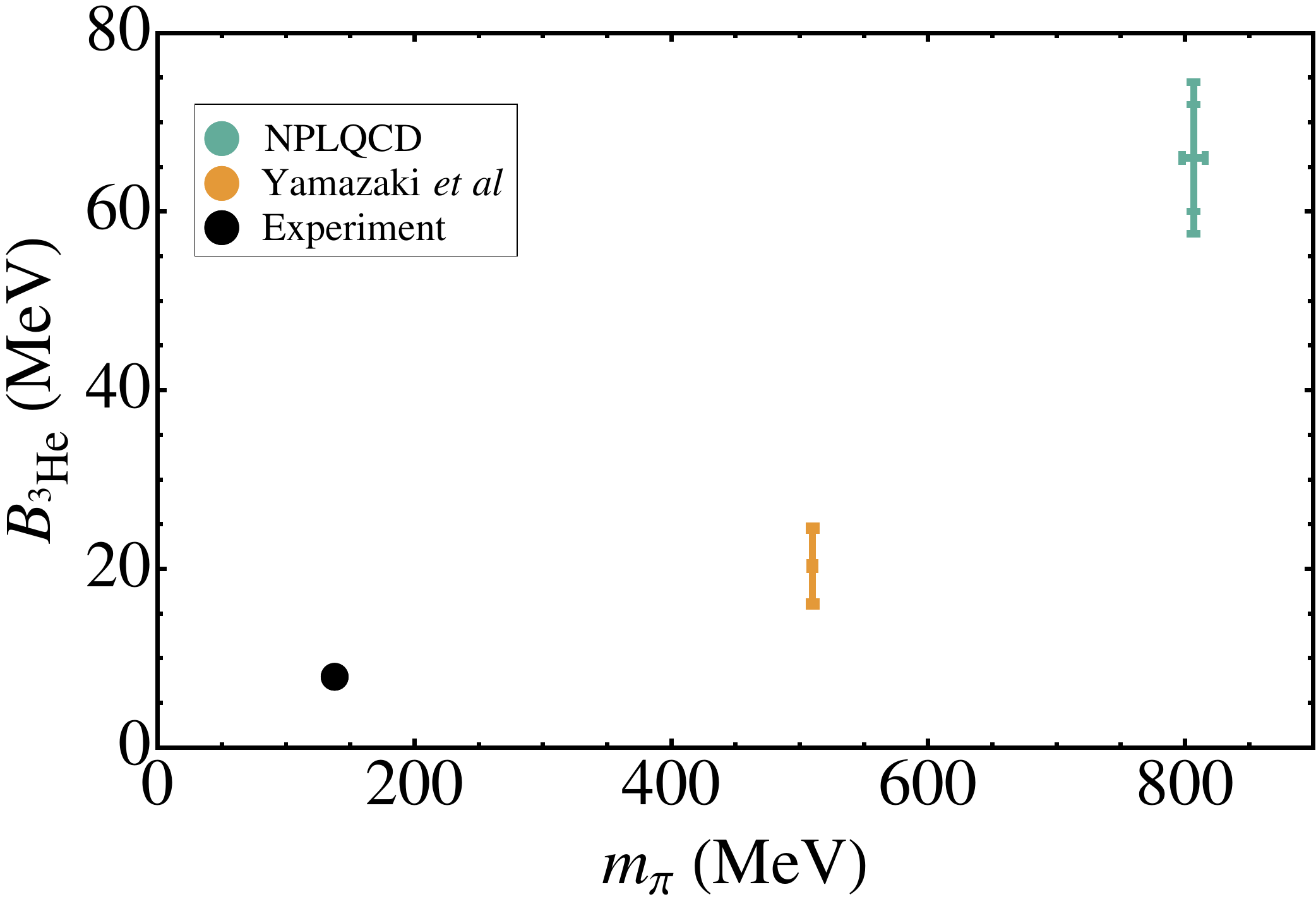}\ \ 
     \includegraphics[width=0.32\textwidth]{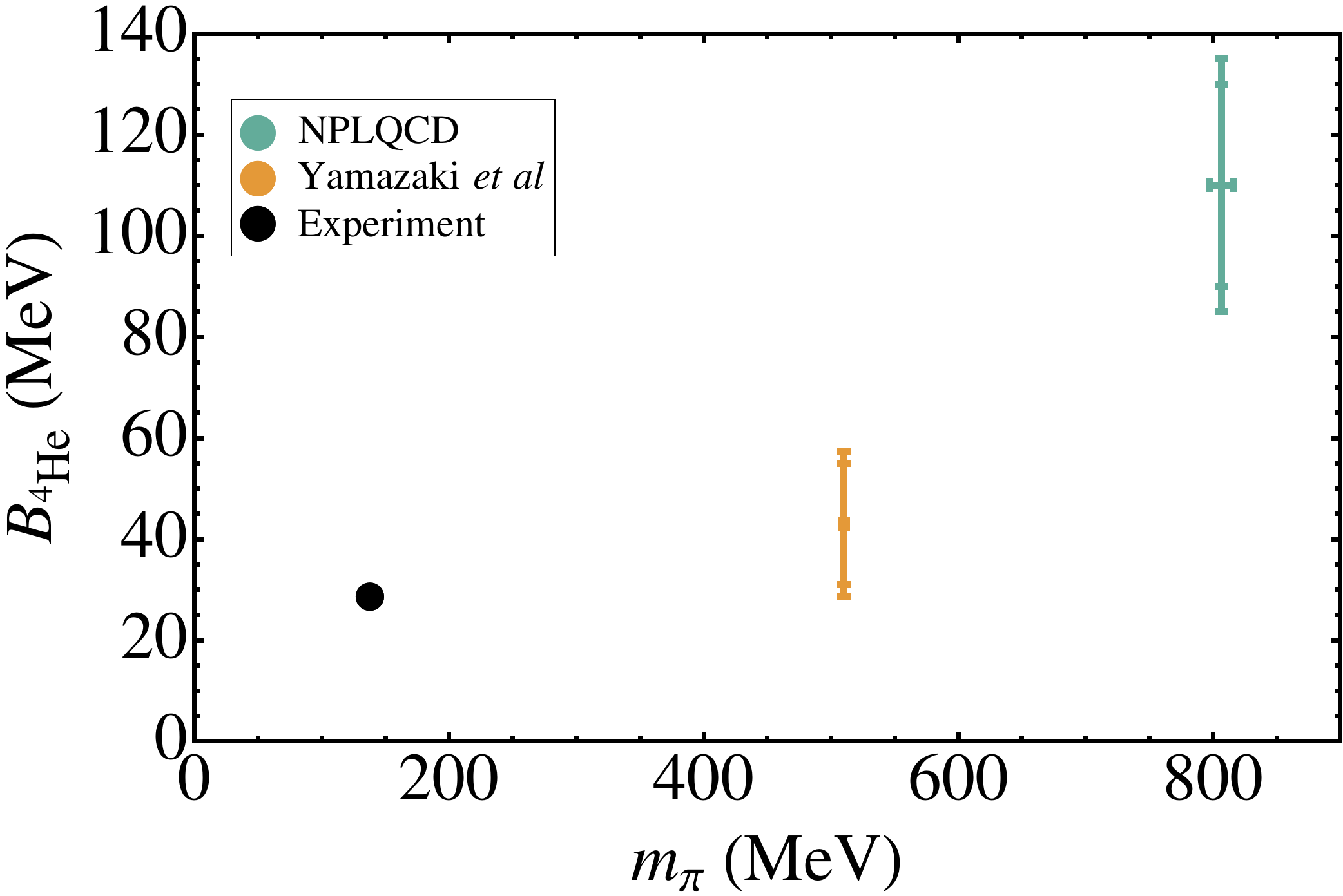}
     \caption{The deuteron (left panel), $^3$He (middle panel) and
       $^4$He (right panel) binding energies from $n_f=2+1$ LQCD
       calculations~\protect\cite{Beane:2011iw,Beane:2012vq,Yamazaki:2012hi}.
       }
  \label{fig:Ball}
\end{figure}

Perhaps some of the most important LQCD calculations of late are those of the
ground states of the light nuclei, including the deuteron, $^3$He, $^4$He and
light hypernuclei.
Fig.~\ref{fig:Ball} shows the binding energy of  the deuteron, $^3$He and 
$^4$He~\protect\cite{Beane:2011iw,Beane:2012vq,Yamazaki:2012hi}
as a function of the pion mass.
Not only is it exciting to see nuclei emerge from QCD for a range of
the light-quark masses, such calculations are crucial in dissecting and refining
the chiral nuclear forces.
However, it is clear that calculations at lighter pion masses are required,
including at the physical pion mass.
\begin{figure}[!ht]
  \centering
  \includegraphics[width=0.7\textwidth]{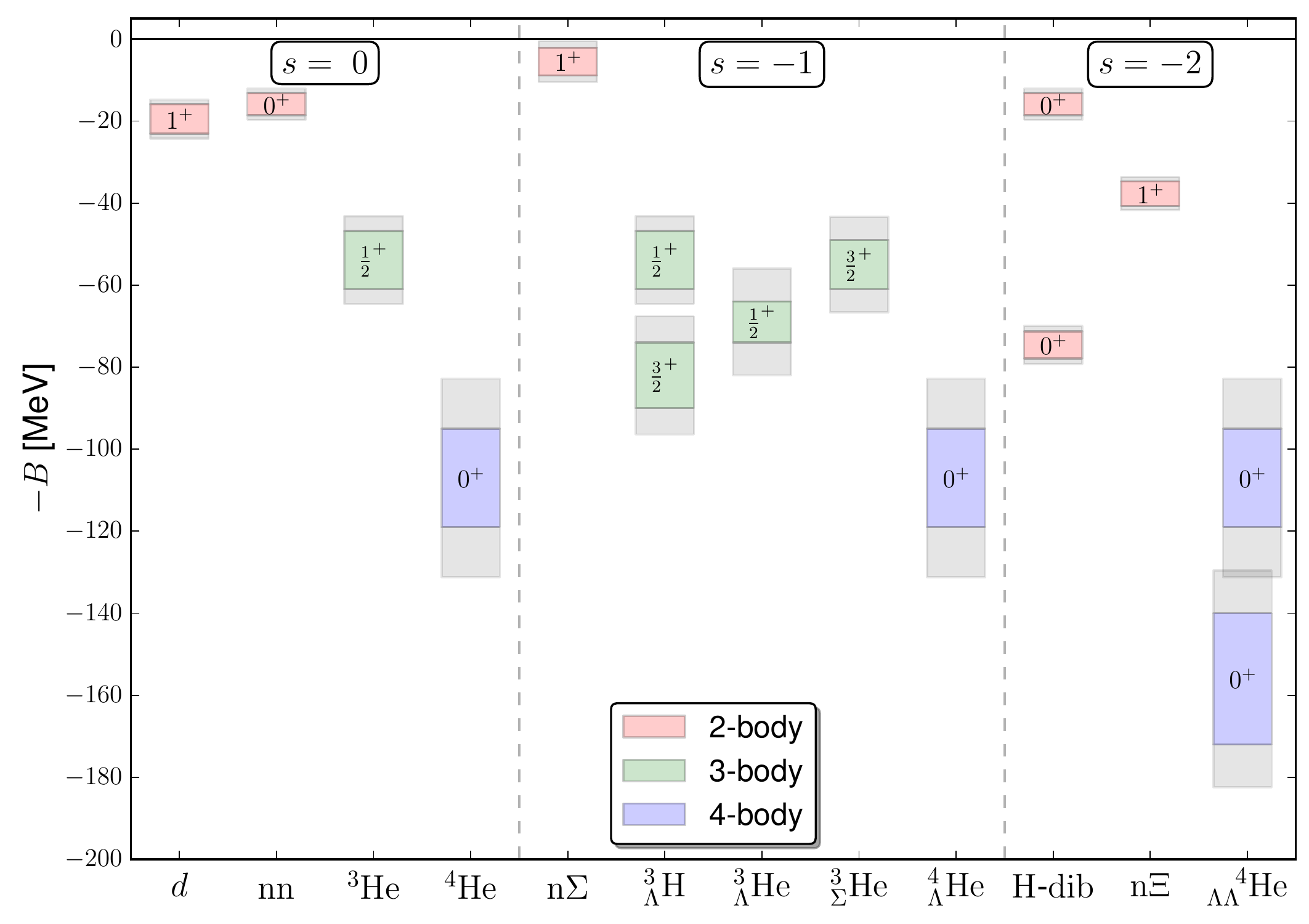}
  \caption{A compilation of the energy levels in light nuclei and hypernuclei
in the limit of flavor SU(3) symmetry
(with spin and parity  $J^\pi$) 
calculated by NPLQCD~\protect\cite{Beane:2012vq} at a pion mass of
$m_\pi\sim 800~{\rm MeV}$.
}
  \label{fig:NuclearSummary}
\end{figure}
A summary of the energy-levels 
at the flavor SU(3) symmetry point
found in the s-shell nuclei and hypernuclei~\protect\cite{Beane:2012vq} is
shown in Fig.~\ref{fig:NuclearSummary}.
These energy levels are elements of SU(3) irreps.  which 
allowed, in some cases, e.g. the H-dibaryon, the hypertriton and $_{\Lambda\Lambda} ^{~\, 4} {\rm He}$,
for distinct  energy levels with the same spin and parity to be determined.
Such calculations will become somewhat more complicated at lighter quark masses
when the up and down
quarks are not degenerate with the strange quark.

The calculations of NPLQCD and those of Yamazaki {\it et al} are already
shedding light on 
how the ground-state energies of the light nuclei approach
their values at the physical light-quark masses.  They are all bound at the
heavier light-quark masses and become less bound as the quarks become lighter.  In the
case of the dineutron, which is bound at $m_\pi\sim 800~{\rm MeV}$, it becomes
unbound at some intermediate value of the pion mass, giving rise to a
neutron-neutron system
with an infinite scattering length.

One of the interesting aspects of the nuclear forces to explore is the tensor
interaction, responsible for the mixing between the S-wave and D-wave channels
in the deuteron channel.
There are a series of LQCD calculations that can be performed that will permit
an extraction of the SD mixing parameter, $\epsilon_1$ using
L\"uscher's method~\cite{Luscher:1986pf,Luscher:1990ux,Briceno:2013lba},
see Ref.~\cite{Briceno:2013bda}.

\subsubsection{Roadblocks of the Past}

It is important to understand how a few of the past roadblocks to progress in
this area have been recently overcome.
One of the roadblocks of the past was/is the ``signal-to-noise problem''
that afflicts states other than the pion.
This problem is seen most simply in the single-nucleon correlation function,
generated with a three-quark source and a three-quark sink.
The variance of this correlation function is dictated by a 3-quark 3-anti-quark
source and a 3-quark 3-anti-quark sink, which 
overlaps with 
both the ${\rm N}\overline{\rm N}$ and $3\pi$ intermediate states (and all
others with the appropriate quantum numbers).
At large times, the variance correlation function is dominated by the $3\pi$ intermediate
state, while the single nucleon correlation function is dominated by the single nucleon,
giving rise to an exponentially degrading signal.  However, at intermediate
times, the behavior of the ``signal-to-noise'' is determined by the overlap of
the variance sinks and sources onto the intermediate hadronic states. 
The
momentum projection
onto single nucleon blocks,  that NPLQCD is currently using, provides a volume suppression
of the $3\pi$ intermediate
state compared to the  ${\rm N}\overline{\rm N}$ state.  
Thus, there is an intermediate  time interval in which the signal-to-noise
ratio is not exponentially degrading.  
It is in this time interval, 
dubbed the ``Golden Window'', that plateaus for the 
low-lying energy levels in light nuclei can be identified.
Unfortunately, the window shrinks as the number of nucleons is increased, and
so further developments will be required to go to much larger nuclei.

A second roadblock that inhibited progress in LQCD calculations of nuclei was
the number of Wick contractions required to form a correlation function.
A system containing $N_u$ up quarks and $N_d$ down quarks requires $N_u ! N_d
!$ Wick contractions, which is a rapidly growing number as one moves beyond the nucleon.
It was recognized that recursion relations relating the Wick contractions
in systems with N mesons can be related to those with $N-1$
mesons~\cite{Detmold:2010au}.
Further, somewhat more sophisticated algorithms~\cite{Doi:2012xd,Detmold:2012eu} have been developed for the
multi-baryon systems that greatly reduce the computing resources required to
perform the contractions. These have led to very efficient calculations of the
s-shell nuclei and hypernuclei, moving beyond the s-shell requires extensions of
these works, and new ideas are required to calculate heavier nuclei.

\subsection{The Bridge Between LQCD and Nuclear Structure}

One of the points of discussion that came up during this presentation was how
to optimally couple the results of LQCD calculations to nuclear structure
calculations.
Given the expertise in the nuclear structure community, it makes little sense
for LQCD theorists to ``go it alone'' and attempt to calculate the entire
periodic table.  
It makes much more sense for the LQCD theorists to produce sets of quantities
that can be handed to the nuclear structure theorists who use them in their
machinery to determine the periodic table.
The question is what are the optimal quantities to pass along from LQCD.

It seems that the minimal set of quantities that could be passed along are the
energy eigenvalues for a given system. 
LQCD calculations 
of the energy spectrum of an A-nucleon system could be performed in
multiple lattice volumes, with multiple lattice spacings and at multiple
light-quark masses, and handed to the the nuclear
structure theorists who in turn reproduce the energies by tuning the chiral
interactions.
These tuned interactions are then used to calculate processes in the continuum.
This methodology was used to calculate the $n\Sigma^-$
interactions at the physical pion mass using $\chi$PT~\cite{Beane:2012ey}.
The chiral interactions were tuned to reproduce the finite-volume energy levels
determined in  a series of LQCD calculations, and then used to calculate the
scattering phase shift at the physical pion mass.
Progress in this direction is starting to be made, as demonstrated in recent
calculations by Nir Barnea and collaborators~\cite{BarneaTalk},
by using the ground state energies of 
the deuteron, dineutron and $^3$He 
at $m_\pi\sim 800~{\rm MeV}$
to reproduce  the $^4$He ground state using the pionless EFT.

\section{Summary and Final Comments}

I have summarized the 
rapid progress that is being made in developing
LQCD into a reliable calculational tool for low-energy nuclear physics.
It holds the promise to directly connect the structure and properties of nuclei
with QCD, and to enable a refinement of the chiral nuclear forces that are used
as input into nuclear structure calculations.
At present, the ground states of the s-shell nuclei and hypernuclei are being calculated
at unphysically heavy light-quark masses, but within the next few
years, such calculations at $m_\pi\sim 140~{\rm MeV}$ will be performed (if
hardware and software resources increase as expected).
Within the next five years, the spectrum and interactions of the lightest
nuclei and hypernuclei
will be postdicted or predicted with fully-quantified uncertainties.

\begin{figure}[!ht]
  \centering
  \includegraphics[width=0.9\textwidth]{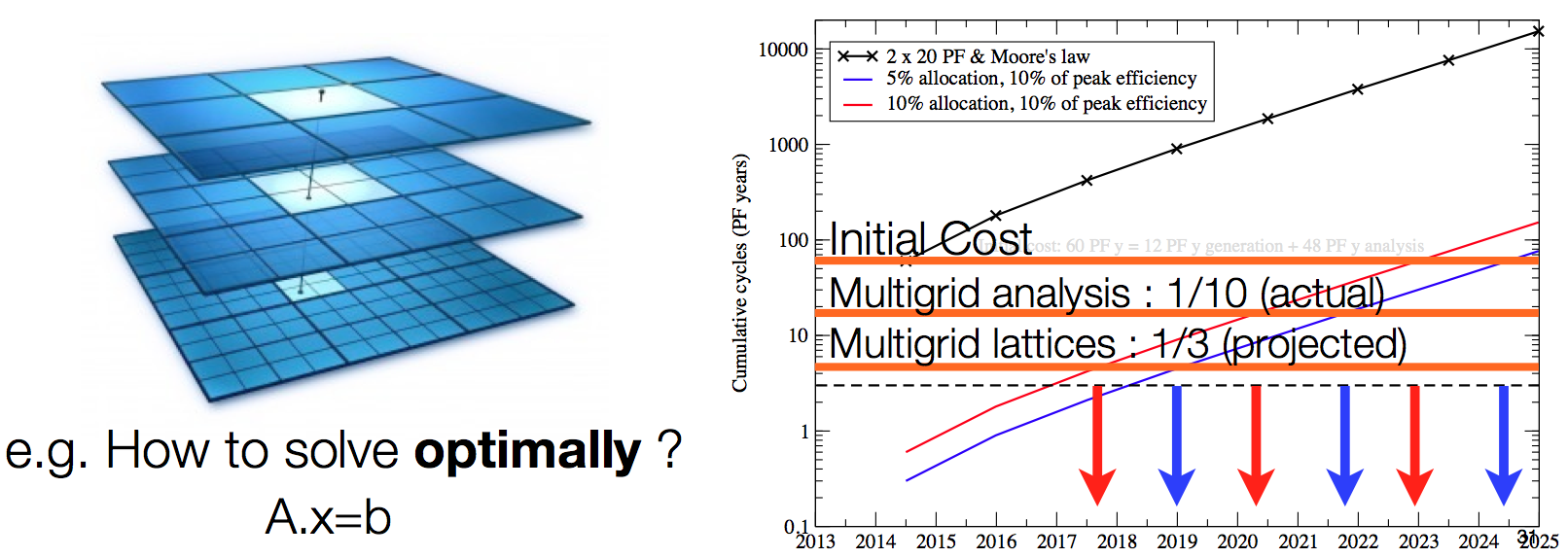}
  \caption{Multigrid  is a recent algorithmic development to be implemented in LQCD
    calculations~\protect\cite{Osborn:2010mb}.
The horizontal (orange) cost estimates 
(that I have added to the original figure) provide one example of what is possible for a
given production scenario.
[Parts of this image~\protect\cite{BJOOtalk} are reproduced with the permission
of B. Joo.]
}
  \label{fig:MGsoftware}
\end{figure}

It is worth emphasizing that the LQCD effort in the US relies heavily on SciDAC
funding to support the scientists who develop and optimize the software to run
on the rapidly evolving computational hardware, e.g. GPU-accelerated compute
nodes that comprise Titan at ORNL, or the BG/Qs at ANL and LLNL.
Further, the effort requires ongoing access to both capability computing
resources on leadership-class  computing facilities, and capacity computing
obtained from NERSC, XSEDE, through USQCD and at local compute clusters.
Ongoing software (see Fig.~\ref{fig:MGsoftware})
and hardware support are critical to progress in this area.

Ideally, one would start with a LQCD calculation and predict all of the
quantities of interest in low-energy nuclear physics.  Presently, we are not in
a position to do this, even if significantly more computing resources were
provided to the program.  
While L\"uscher provided the formalism to relate the
two-body S-matrix directly to two-particle energy levels inside a cubic volume
with the fields subject to periodic boundary conditions~\cite{Luscher:1986pf,Luscher:1990ux}, which has since been
understood and generalized to the two-nucleon systems, e.g. Ref.~\cite{Briceno:2013lba}, 
such formalism is complicated to apply in coupled-channels systems~\cite{Hansen:2012tf,Briceno:2012rv,Guo:2012hv}.
Further, the  formalism is not in place for the three- and higher-body 
sectors, but progress is being  made in such systems~\cite{Briceno:2012yi,Polejaeva:2012ut}.

In closing, great progress is being made to reliably determine and refine the
nuclear forces directly from QCD using Lattice QCD.

\vskip 0.1in

{\it {\bf Happy Birthday James}: 
James Vary is one of the first nuclear theorists I met when I arrived in the United States
to enter the PhD program at Caltech in the mid 1980's. I recall James taking
the time to talk physics with me during his stay.  
His detailed knowledge of, and passion for, 
important problems  of the day left a lasting impression on me.
Despite having been able to chat  with, and even collaborate with,  James since that time, 
when I learned that this conference was in part to celebrate James's 70th
birthday, I was taken aback as it seems like yesterday that he was in his early 40's
(and I was in my early 20's), and he has retained the same passion and energy
for science.  I should also add that James is responsible for me remembering
the value of $\hbar c$!
Happy 70$^{\rm th}$!!
}

\end{document}